\documentclass[prb,showpacs,twocolumn,superscriptaddress,aps]{revtex4-1}
\usepackage[utf8]{inputenc}
\usepackage{graphicx, xcolor}
\usepackage{dcolumn}
\usepackage{bm}
\usepackage{amsmath,amsthm,amssymb,bbold}
\usepackage{color}
\usepackage{verbatim}
\usepackage{ulem}
\usepackage{nicefrac}
\usepackage{float}
\usepackage{natbib}

\usepackage{physics}

\usepackage{lipsum} % placeholder text

\usepackage{bbold}
\newcommand{\id}{\ensuremath{\mathbb{1}}}

\usepackage{bm}  % provides \boldsymbol
\renewcommand{\vec}[1]{\boldsymbol{\mathbf{#1}}}
\newcommand{\unitvec}[1]{\hat{\boldsymbol{\mathbf{#1}}}}

\definecolor{darkGreen}{RGB}{0,110,0}
\definecolor{darkBlue}{RGB}{0,0,130}
\usepackage[colorlinks,citecolor=darkGreen,linkcolor=darkBlue,urlcolor=blue,hyperindex]{hyperref}

\def\equationautorefname~#1\null{Eq. (#1)\null}

\newcommand{\appref}[1]{\hyperref[#1]{App.~\ref*{#1}}}

\begin{document}

\title{Hierarchy of energy scales and field-tunable order by disorder in dipolar-octupolar pyrochlores}

\author{Benedikt Placke}
\author{Roderich Moessner}
\author{Owen Benton}
\affiliation{Max-Planck-Institut f\"ur Physik komplexer Systeme, N\"othnitzer Stra\ss{}e 38, 01187 Dresden, Germany}

\date{\today}

\begin{abstract}
Dipolar-octupolar pyrochlore magnets in a strong external magnet field applied in the [110] direction are known to form a `chain' state, with subextensive degeneracy. Magnetic moments are correlated along one-dimensional chains carrying effective Ising degrees of freedom which are noninteracting on the mean-field level.
Here, we investigate this phenomenon in detail, including the effects of quantum fluctuations.
We identify two distinct types of chain phases, both featuring distinct subextensive, classical ground state degeneracy. 
Focussing on one of the two kinds, we discuss lifting of the classical degeneracy by quantum fluctuations. We map out the ground-state phase diagram as a function of the exchange couplings, using linear spin wave theory and real-space perturbation theory. 
We find a hierarchy of energy scales in the ground state selection, with the effective dimensionality of the system varying in an intricate way as the hierarchy is descended. We derive an effective two-dimensional anisotropic triangular lattice Ising model with only three free parameters which accounts for the observed behavior. 
Connecting our results to experiment, they are consistent with the observation of a disordered chain state in Nd$_2$Zr$_2$O$_7$. We also show that the presence of two distinct types of chain phases has consequences for the field-induced breakdown of the apparent $U(1)$ octupolar quantum liquid phase recently observed in Ce$_2$Sn$_2$O$_7$.
\end{abstract}

\maketitle

\section{Introduction}

Frustrated magnets consisting of corner-sharing fully connected units often feature nontrivial ground state degeneracy \cite{Moessner1998}.
This is captured by local constraints on the configuration of each unit, which nevertheless leave the system with a large number of unconstrained degrees of freedom. In the case of the pyrochlore lattice, which consists of corner-sharing tetrahedra, this often takes the form of an ice-like two-in-two-out rule, leading to the class of materials known as ``spin-ice'' \cite{Harris1997, Bramwell2001}.

One natural question to ask is what is the fate of this classical degeneracy in the presence of quantum fluctuations. Since the degeneracy is ``accidental'' in the sense that the degenerate states are not related by a symmetry of the Hamiltonian, it is expected to be lifted by fluctuations, a process which is generally called order by disorder (OBD) \cite{villain1980, henley1987, henley1989}. Probably the most prominent experimental example of quantum OBD is the effective spin-$\tfrac{1}{2}$ pyrochlore Er$_2$Ti$_2$O$_7$ \cite{ Poole2007, Ruff2008, Champion2003, Zhitomirsky2012, Savary2012, Ross2014, Rau2016}. 
In Er$_2$Ti$_2$O$_7$ the OBD mechanism lifts the degeneracy of a manifold of antiferromagnetic states with a single $U(1)$ degree of freedom. There are competing contributions to the ground state selection \cite{Rau2016, rau2019} which cannot be easily tuned, making them difficult to disentangle from one another.

In this article, we consider a different, experimentally motivated, case of OBD on the pyrochlore lattice: the case of dipolar-octupolar pyrochlores in a [110] magnetic field.
Here, in contrast to Er$_2$Ti$_2$O$_7$, the mean field degeneracy grows sub-extensively.
Both the energy scale of ground state selection, and the selected ground state can be tuned by an external field, as also seen in some theoretical models on other lattices \cite{Hassan2006, Henry2014}.

\begin{figure}
    \centering
    \includegraphics{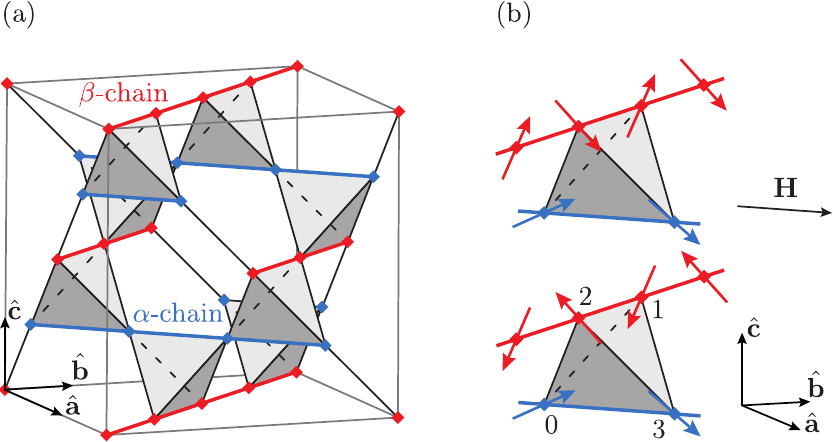}
    \caption{(a) Pyrochlore lattice separated into chains parallel ($\alpha$ chains) and perpendicular ($\beta$ chains) to the external field $\vec H \parallel (1,1,0)$. (b) The two classical ground state orientations for a single tetrahedron with the local easy axis along the local [111] direction. While the magnetic moments on the $\alpha$ chains are pinned by the external field $\vec h$, those on the $\beta$ chains are perpendicular to the field and thus each retain one independent Ising-like degree of freedom. On one of the tetrahedra, we also indicate the four fcc sublattices $0, 1, 2, 3$. The local magnetic moments on each sublattice point into a direction $\pm\unitvec z_i$, where the $\unitvec z_i$ are defined to point into the tetrahedron.}
    \label{fig:geom}
\end{figure}

Dipolar-octupolar pyrochlores are a family of rare-earth oxides R$_2$M$_2$O$_7$ with R=Nd, Ce, Sm \cite{Bertin2015, Hallas2015, Lhotel2015,Xu2016, Malkin2010, Singh2008, Pecanha2019, Gao2019, Gaudet2019}.
They are well described by a nearest-neighbor XYZ model with a peculiar coupling to an external magnetic field \cite{Huang2014}. The dipolar-octupolar crystal field doublet couples only to the projection of the external field onto the local [111] easy axes, which are not collinear. A uniform field applied in the crystalline direction [110] thus separates the lattice into two sets of one-dimensional chains (see \autoref{fig:geom}), with a frustrated interaction that cancels out exactly on the mean-field level. 

Here, we investigate this phenomenon through calculations of the classical phase diagram and by showing how quantum fluctuations lift the degeneracy of the chain states and restore correlations between chains, albeit at extremely small energy scales.

Studying the classical phase diagram as a function of exchange parameters and field, we find four distinct chain phases, which can be divided into two different types. First, there are three ${\rm Chain}_\lambda$ ($\lambda = x, y, z$) phases, in which the chains along the field direction ($\alpha$ chains) are polarized, while those perpendicular to it ($\beta$ chains) each retain one Ising-like degree of freedom [see \autoref{fig:geom}]. This yields a subextensive ground-state entropy scaling as the square of the linear system size $L^2$ when $N\sim L^3$. 

There is then one further type of chain phase, ${\rm Chain}_Y^*$ which appears due to the special status of the pseudospin $y$-axis in the XYZ model for dipolar octupolar pyrochlores. The operator $S^y_i$ transforms like a magnetic octupole \cite{Huang2014} and therefore does not couple linearly to an external field. This gives rise to the ${\rm Chain}_Y^*$ phase at low fields if the octupolar exchange coupling $J_y$ is dominant.
At low fields, the moments on the $\alpha$ chains  have a finite octupolar component, which results in an additional independent Ising degree of freedom per $\alpha$ chain.
The existence of this additional chain phase for strong $J_y$ implies the possibility of a field-induced confinement transition out of the octupolar quantum liquid state recently proposed for Ce$_2$Sn$_2$O$_7$\cite{Sibille2020}.

Turning to the effect of quantum fluctuations, we show that for
the ${\rm Chain}_\lambda$ phases, fluctuations via the polarized $\alpha$ chains mediate an effective interaction between the unpolarized $\beta$ chains and thus lift the subextensive classical degeneracy. We compute the leading effective chain interactions and map out the ground-state phase diagram in the semiclassical limit $S\to\infty$. 
To this end, we use real space perturbation theory (RSPT) \cite{Long1989, Heinila1993, Chernyshev2014, Rousochatzakis2017} to derive a two-dimensional model describing the effective interaction between the chains, which takes the form of a nearest-neighbor anisotropic triangular lattice Ising model.
To corroborate our results, we also compute the zero point energy of different classical ground state spin configurations numerically using linear spin wave theory (LSWT) and find that the full splitting of the classically degenerate states can be captured by the effective model remarkably well. The two methods agree also on a quantitative level, with positions of phase boundaries agreeing even for moderately strong transverse couplings.

Two kinds of ground state are selected by OBD, depending on the exchange parameters and external field. First, a completely uniform state, where all $\beta$ chains align ferromagnetically and second, zigzag configurations where $\beta$ chains align antiferromagnetically in the [110] direction.
The selection is governed by two competing effective nearest-neighbor chain interactions generated by correlated fluctuations around hexagons. The two couplings scale differently as a function of the external magnetic field, making it possible, in some parts of the phase diagram, to tune between the ferromagnetic and zigzag ground states.
The left-over degeneracy in the zigzag phase is lifted on a minuscule energy scale that however can be resolved in our LSWT calculation. In terms of the effective model, the splitting can be fully understood by including an effective next-nearest-neighbor coupling, which appears at higher order in RSPT.

Discussing the effect of temperature we compute the classical low-temperature expansion around different ground states. Remarkably, we find that, in contrast to its quantum counterpart, at leading order classical OBD is completely absent in all ${\rm Chain}_\lambda$ phases.

Guruciaga {\it et al.}\cite{Guruciaga2016} have considered a related problem of thermal OBD in a classical pyrochlore Ising model in a [110] field.
The case considered here differs from [\onlinecite{Guruciaga2016}] in that we consider quantum OBD with a Hamiltonian directly motivated by experiments on dipolar-octupolar pyrochlores\cite{Xu2019, Lhotel2018, Petit2016, Benton2016}, and in that we find multiple different ground states as a function of field and exchange parameters.

The remainder of this paper is organized as follows. In \autoref{sec:dipolar-octupolar}, we review dipolar-octupolar pyrochlores and their classical ground-state phase diagram in a [110] field.
This serves as a starting point for the OBD calculation presented in the following. In \autoref{sec:eff-Ising}, we derive a two-dimensional triangular lattice Ising model describing the lifting of the classical degeneracy in leading-order real space perturbation theory (RSPT). In \autoref{sec:obd-phases}, we compute the ground-state phase diagram using linear spin wave theory (LSWT), comparing the
results with those from RSPT.
In \autoref{sec:field} we discuss the tunability of the ground state selection as a function of the external field strength. We compare the results of the LSWT calculation with those from RSPT. Section \ref{sec:experiment} explores the connection of our work to experiment. We conclude in \autoref{sec:conclusion}.

\section{Dipolar-Octupolar Pyrochlores\label{sec:dipolar-octupolar}}

\subsection{Exchange Hamiltonian}

When describing the low-energy properties of rare-earth magnets, we have to start with the the local crystal electric
field (CEF). The CEF energy scales are typically orders of magnitude larger than the exchange energy between two ions \cite{rau2019}. To calculate low temperature, low energy properties, it is therefore usually sufficient to consider the lowest energy multiplet of the crystal field. For materials within our present interest, this takes the form of a doublet.

On the pyrochlore lattice [\autoref{fig:geom}] this doublet comes in three different flavors.
Each flavor corresponds to an irreducible representation of the double group of the point group D$_{\mathrm{3d}}$, together with time reversal invariance. In the case that the two states forming the lowest CEF are not related by spatial symmetry but only by time-reversal, it is called a dipolar-octupolar doublet \cite{Huang2014}. This occurs in rare-earth pyrochlores based on Nd, Ce and Sm.

Once projected into a dipolar-octupolar doublet, the only surviving component of the magnetic moment $\vec \mu_i $ operator is that oriented along the local [111] easy axis $\unitvec z_i$, which connects the centers of the two tetrahedra sharing site $i$. In our convention, the $\unitvec z_i$ are constant on each of the four sublattices $\mathcal L_{\nu}$ ($\nu=0, 1, 2, 3$) and point into the tetrahedron shown in figure \autoref{fig:geom} (b).

Defining pseudospin-$\tfrac{1}{2}$ operators $S^{\alpha}_i$ acting within the doublet, one can choose a basis such that
\begin{eqnarray}
\vec \mu_i = \mu_{\mathrm{B}} g_z \unitvec z_i S_i^z.
\end{eqnarray}

While $S^{x}_i$ and $S^{z}_i$ transform like the $\unitvec z_i$ component of a magnetic dipole moment, $S^{y}_i$ transforms like a component of a magnetic octupole tensor \cite{Huang2014}. Expressed in terms of components of the angular momentum operator $\mathcal J$, the corresponding octupole moment is \cite{Patri2020, Sibille2020}
\begin{equation}
    S^{y} \propto \mathcal J_y^3 - \mathcal J_y \mathcal J_x^2 - \mathcal J_x \mathcal J_y \mathcal J_x - \mathcal J_x^2 \mathcal J_y,
    \label{eq:octupole}
\end{equation}
which transforms trivially under all lattice symmetries (D$_{\mathrm{3d}}$) but is odd under time reversal.

The most general symmetry allowed nearest-neighbor exchange Hamiltonian for such a dipolar-octupolar pyrochlore magnet then only couples pseudospin components with the same transformation properties
\begin{subequations}
\begin{align}
    \mathcal H =& \mathcal H_{\mathrm{ex}} + \mathcal H_{\mathrm{field}}, \\
    \mathcal H_{\mathrm{ex}} =& \sum_{\expval{ij}} J_{xx} S_i^x S_j^x + J_{yy} S_i^y S_j^y + J_{zz} S_i^z S_j^z \nonumber\\
        &\phantom{\sum_{\expval{ij}}}+ J_{xz} \left( S_i^x S_j^z + S_i^z S_j^x \right),\\
    \mathcal H_{\mathrm{field}} =&  -\mu_{B} g_z \sum_i \vec H \cdot \unitvec z_i S_i^z,
\end{align}
\end{subequations}
where the first sum in the exchange Hamiltonian $\mathcal H_{\rm ex}$ runs over all nearest-neighbor pairs $\expval{ij}$ in the pyrochlore lattice and the sum in the field Hamiltonian $\mathcal H_{\rm field}$ runs over all sites $i$.

We can remove the $XZ$ exchange by a rotation of the pseudospin axes by an angle $\theta$ around the pseudospin $y$-axis
\begin{subequations}
\begin{align}
    S_i^x &\to \cos\theta\,S_i^x - \sin\theta\,S_i^z, \\
    S_i^z &\to \sin\theta\,S_i^x + \cos\theta\,S_i^z, \\ 
    \tan(2\theta) &= \frac{2 J_{xz}^{(0)}}{J_{xx}^{(0)} - J_{zz}^{(0)}},
\end{align}
\end{subequations}
which yields
\begin{subequations}
\begin{align}
    \mathcal H &= \mathcal H_{\mathrm{ex}} + \mathcal H_{\mathrm{field}}, \\
    \mathcal H_{\mathrm{ex}} &= \sum_{\expval{ij}} J_x S_i^x S_j^x + J_y S_i^y S_j^y + J_z S_i^z S_j^z, \\
    \mathcal H_{\mathrm{field}} &=  - \mu_{\mathrm{B}} g_z \sum_i \vec H \cdot \unitvec z_i \left(\sin\theta\, S_i^x + \cos\theta\, S_i^z\right).
\end{align}
\label{eq:hamiltonian}
\end{subequations}
To avoid too ornamented notation, we have not renamed the spin operators $S_i^\alpha$. From here on, the $S_i^\alpha$ refer to those in \autoref{eq:hamiltonian}.
Altogether, dipolar-octupolar pyrochlores are described by a relatively simple XYZ exchange Hamiltonian, however with a peculiar coupling to the magnetic field.

\subsection{Classical phase diagram in a [110] field \label{sec:classical-phases}}

\begin{figure*}
    \centering
    \includegraphics{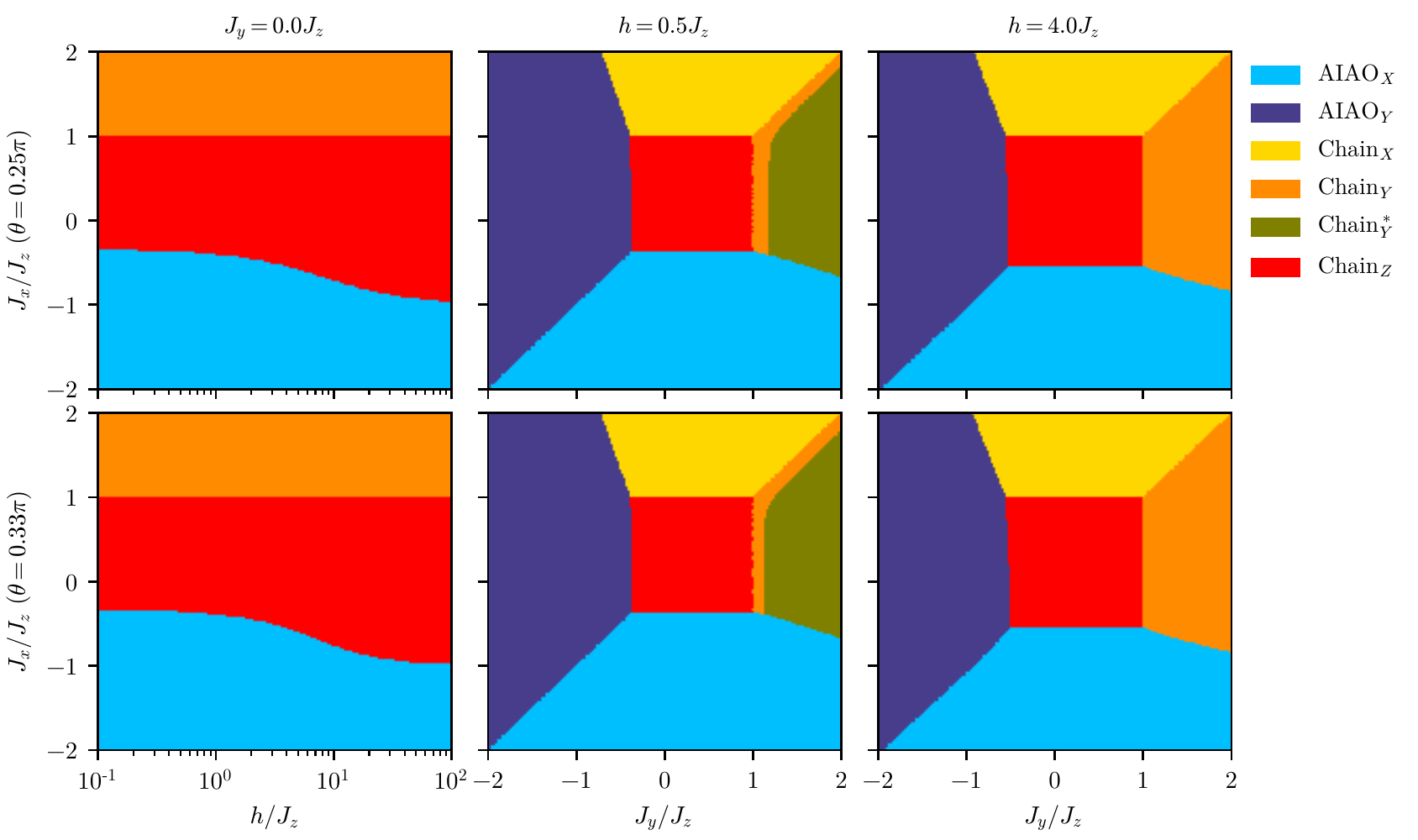}
    \caption{Zero-temperature mean-field phase diagram of dipolar-octupolar pyrochlores
    with an applied field along the [110] direction as a function of exchange couplings and field strength, for $J_z>0$. The first column is for fixed octupolar coupling $J_y=0$ and the second and third coupling are for fixed magnetic fields $h=J_z/2$ and $h=4J_z$, respectively. The top and bottom rows have a fixed mixing angle $\theta=\pi/4$ and $\theta=0.33\pi$ respectively (cf. \autoref{eq:hamiltonian} for the meaning of the parameters). 
    The phase diagram does not change qualitatively for any value of $\theta$ except for the singular limits $\theta\to0$, $\theta\to\pi/2$, and $\theta\to\pi$. As $h\to0$ (not shown), while the low-field all-in-all-out (${\rm AIAO}_{\lambda}$) phases are stable, the chain phases (${\rm Chain}_{\lambda}^{(*)}$) are not. In the dipolar case ($\lambda=x,z$), at exactly zero field the chain phases become the respective spin-ice phase (${\rm SI}_{\lambda}$), with extensive instead of subextensive degeneracy. 
    For the octupolar chain phase, as the field is lowered starting in the ${\rm Chain}_Y$ phase, there is actually a second-order transition to an intermediate phase, ${\rm Chain}_Y^*$ [see also \autoref{fig:chain-y-h}], in which each $\alpha$ chain carries an additional Ising degree of freedom. The ${\rm Chain}_Y^*$ phase is  connected to the octupolar ice phase ${\rm SI}_Y$ at zero field.
    At large fields, the orientation of magnetic moments in the ${\rm AIAO}_{\lambda}$ phases get significantly distorted with respect to their zero field orientations. However, even as $h\to\infty$ they retain their two-fold degeneracy and nonzero order parameter ($\sum_i S_i^\lambda \geq SN/2$).}
    \label{fig:classical-phases}
\end{figure*}

Before turning to the case of a finite field, we quickly discuss the zero-field phase diagram. For zero field ($\vec H=0$) there are six possible classical ground state phases. 

There are the three ice-like phases ${\rm SI}_{\lambda} ~ (\lambda=x,y,z)$, stabilized when the
corresponding exchange coefficient is positive and sufficiently strong
\begin{equation}
    J_{\lambda} > \max(-3J_{\lambda'}, J_{\lambda'}) ~~ \forall \lambda' \neq \lambda.
\end{equation}
In these phases, all pseudospins align with the $\lambda$ axis and are governed by a local ``two up, two down'' constraint with respect to this axis on every tetrahedron. This rule leaves an extensive number of degrees of freedom. If $\lambda=x,z$ then the rule corresponds to
a ``two-in-two-out" rule on the magnetic moments; i.e. for each tetrahedron, two of the four magnetic moments on its corners point into and two point out of it. ${\rm SI}_{x}$ and ${\rm SI}_{z}$ thus describe the well known spin ice state, whereas ${\rm SI}_{y}$ describes a state analogous to spin ice but where the active degrees of freedom are octupolar moments [\autoref{eq:octupole}].

There are then three ordered, ``all-in-all-out'' phases (${\rm AIAO}_{\lambda}, ~ \lambda=x,y,z$), occurring for sufficiently strong negative exchange coefficient
\begin{equation}
    J_{\lambda} < \min(-J_{\lambda'}/3, J_{\lambda'}) ~~ \forall \lambda' \neq \lambda.
\end{equation}
These correspond to ferromagnetic order of the pseudospins along the $\lambda$ axis. If $\lambda=x,z$ then this will correspond to antiferromagnetic all-in-all-out order of the magnetic moments. The ${\rm AIAO}_y$ phase, by constrast, is a form of octupolar order.

Turning to finite fields, for the rest of this article we set 
\begin{equation}
    \vec H = \frac{\sqrt 3 h}{2\mu_{\mathrm{B}} g_z} \mqty(1\\ 1\\ 0).
    \label{eq:field}
\end{equation}
Substituting this into \autoref{eq:hamiltonian} yields:
\begin{align}
    \mathcal H_{\mathrm{field}} =&  - h \sum_{i\in \mathcal L_0} \left(\sin\theta\, S_i^x + \cos\theta\, S_i^z\right) \nonumber\\
        &+h \sum_{j\in \mathcal L_3} \left(\sin\theta\, S_j^x + \cos\theta\, S_j^z\right),
        \label{eq:hamiltonian-field}
\end{align}
where $\mathcal L_\nu ~ (\nu=0,1,2,3)$ denotes the four fcc sublattices, also indicated in \autoref{fig:geom} (b). The field, when applied in this direction, only couples to sites on the sublattices $0$ and $3$. As shown in \autoref{fig:geom} (a), sites on which the magnetic moment couples to the external field then form one-dimensional chains which are parallel to the field direction ($\alpha$ chains), whereas sites whose magnetic moment does not couple to the external field also form one-dimensional chains, which are perpendicular to the field direction ($\beta$ chains). 
%Note that for all sites, nearest neighbors either lie on the same chain or on a chain of the opposite type.

The classical ground state phase diagram of dipolar-octupolar pyrochlores with ${\bf H}\parallel(1,1,0)$ is shown in  \autoref{fig:classical-phases}. There are now seven possible phases, six of which are directly connected to the six phases possible at zero field. The ${\rm SI}_Y$ phase evolves into two distinct chain phases as a function of field strength, separated by a second-order transition. 

At any finite field the extensive degeneracy of the three spin ice phases ${\rm SI}_{\lambda}$ is broken down to a subextensive degeneracy, yielding effectively one-dimensional degrees of freedom.
We first consider the dipolar cases, i.e. $\lambda=x,z$, shown in \autoref{fig:geom} (b): while the magnetic moments on the $\alpha$ chains (sublattices $0$ and $3$) are polarized by the field, magnetic moments on the $\beta$ chains do not couple to the field at all, but are restricted to two possible configurations per $\beta$ chain by the two-in-two-out rule. Thus, each $\beta$ chain carries an independent effective Ising degree of freedom. We denote this phase when resulting from dimensional reduction in a ${\rm SI}_{\lambda}$ phase by ${\rm Chain}_{\lambda}$. In all of them, there is a finite magnetization in the direction of the field, stemming from the polarized $\alpha$ chains and each $\beta$ chain carries a finite magnetization in the direction of the chain multiplied by its respective Ising degree of freedom.

The case of the octupolar spin ice phase ${\rm SI}_Y$ is special since the ordered moment does not couple to the field. Because of this, the low-field phase which evolves out of the octupolar ice ${\rm SI}_Y$ is distinct from the other chain phases.
If the octupolar exchange $J_y$ dominates, a weak field partially polarizes the magnetic moments on the $\alpha$ chains, but leaves a finite octupolar component $S_i^y$. This component does not couple to the external field even on the $\alpha$ chains, however the ``spin-ice'' rule $\sum_{\mathrm{tet}} S^y_i = 0$ enforces ordering of those moments along both $\alpha$ and $\beta$ chains. 
Hence, again the extensive degeneracy is broken down to a subextensive degeneracy, but with an Ising degree of freedom residing on each $\alpha$ and each $\beta$ chain. We call this phase ${\rm Chain}_Y^*$.

\begin{figure}
    \centering
    \includegraphics{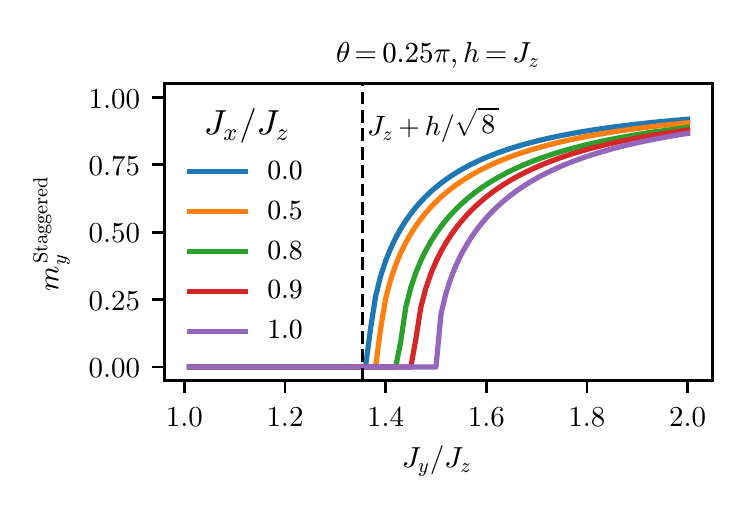}
    \caption{Transition form the ${\rm Chain}_Y$ to the ${\rm Chain}_Y^*$ phase as function of $J_y$ for fixed field $h=J_z$ and $\theta=\pi/4$. The plot shows the staggered octupolar moment on the $\alpha$ chains [\autoref{eq:staggered-m}], which is nonzero only in the ${\rm Chain}_Y^*$ phase.}
    \label{fig:chain-y}
\end{figure}

The additional Ising degrees of freedom in the ${\rm Chain}_Y^*$ phase are related to the staggered octupolar moment on the $\alpha$ chains, which on a chain $C$ is given by
\begin{equation}
   m_{y}^{\rm Staggered} = \frac{1}{2S} \left(\sum_{i \in \mathcal L_0 \cap C } S_i^y -\sum_{j \in \mathcal L_3 \cap C} S_j^y\right).
    \label{eq:staggered-m}
\end{equation}
As the field $h$ is increased, or the octupolar exchange coupling $J_y$ is lowered, this quantity vanishes in a second order transition [\autoref{fig:chain-y} and \autoref{fig:chain-y-h}] marking the onset of a chain phase ${\rm Chain}_Y$ which is equivalent to the dipolar chain phases ${\rm Chain}_{X/Z}$ in that there is exactly one Ising degree of freedom per $\beta$ chain.

The three ${\rm AIAO}_{\lambda}$ phases are stable in a small magnetic field. However, if the respective coupling $J_\lambda$ is sufficiently weak (i.e. if  $|J_\lambda| < {J_{\lambda'}}~ {\rm for \ some \ } \lambda' \neq \lambda$), then the order is destroyed at a finite field, marking a transition to a ${\rm Chain}_{\lambda'}$ phase.
Note that in a finite field, the ${\rm AIAO}_{\lambda}$ states will acquire a finite magnetization since the magnetic moments on the $\alpha$ chains will be partially polarized. However, even as $h\to\infty$, the phase retains an intensive two-fold degeneracy and a nonzero order parameter $\sum_i S_i^\lambda \geq SN/2$. These two ${\rm AIAO}$ states are related by a mirror symmetry in the case of the dipolar ${\rm AIAO}_{X}$ and ${\rm AIAO}_{Z}$ phases and by a combination of time-reversal and a mirror symmetry in the case of the octupolar ${\rm AIAO}_{Y}$ phase.

\section{Effective triangular lattice Ising model\label{sec:eff-Ising}}

\begin{figure}
    \centering
    \includegraphics{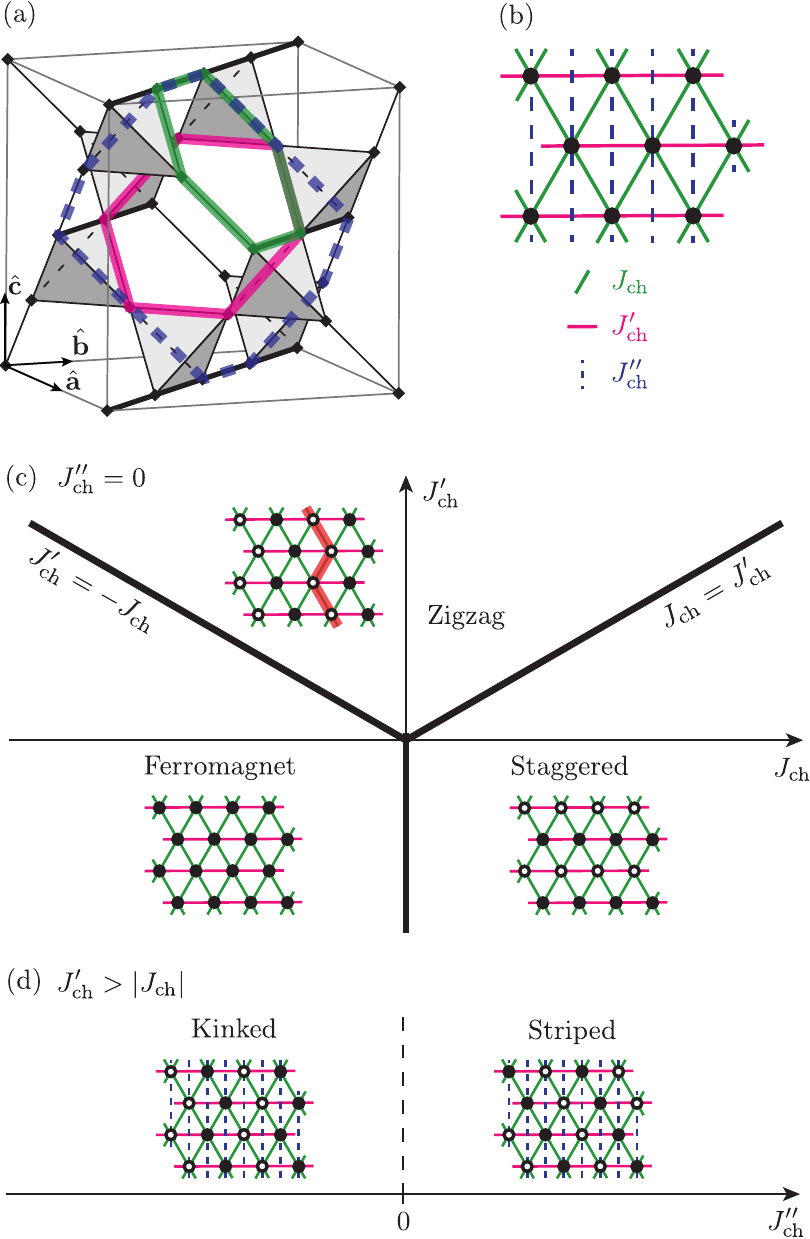}
    \caption{Effective triangular lattice Ising model. (a) the smallest linked clusters that yield a correction to the classical energy and distinguish between classical ground states. The part of the energy correction that distinguishes between classical ground states is proportional to the product of the Ising variables on the chains they connect. (b) the resulting effective anisotropic triangular lattice Ising model. (c) the phase diagram of the anisotropic triangular lattice Ising model as a function of the two nearest-neighbors couplings for the case of negligible next-nearest-neighbor coupling\cite{Dublenych2013}. (d) lifting of the subextensive degeneracy in the zigzag phase by next-nearest-neighbor coupling $J_{\mathrm{ch}}''$.}
    \label{fig:ising}
\end{figure}

As established in the previous section, if none of the exchange constants $J_{\lambda}$ are strongly negative, the system at moderate fields is in a chain phase [\autoref{fig:classical-phases}], where the classical ground state is subextensively degenerate with an Ising degree of freedom carried by each $\beta$ chain [\autoref{fig:geom}]. The interaction between those Ising degrees of freedom is frustrated and cancels out exactly on the mean-field level. However, since the resulting classical ground state degeneracy is accidental in the sense that the degenerate states are not related by any symmetry of the Hamiltonian [\autoref{eq:hamiltonian}] one would expect it to be lifted by quantum fluctuations, a process called (quantum) order-by-disorder (OBD).

\subsection{Nearest-neighbor model}

In this section, we derive the leading order contribution to the effective interaction between the chains using real space perturbation theory (RSPT) \cite{Long1989, Heinila1993, Chernyshev2014, Rousochatzakis2017}. The method divides the Hamiltonian into an unperturbed part, which has a particular classical state as its ground state, and a perturbation incorporating all transverse couplings. One then performs standard perturbation theory to obtain the corrections to the classical energy due to the transverse terms. Since the perturbation depends parametrically on the explicit classical state, the energy correction will depend on it too, possibly lifting classical degeneracies.

We restrict the discussion to the ``simple'' chain phases ${\rm Chain}_\lambda$ where the effective interactions take the form of a nearest-neighbor anisotropic triangular-lattice Ising model.

We can parameterize any classical ground state $\{\vec S_i^{(0)}\}$ by $L^2$ Ising variables $\{\eta_i\}$, $\eta_i \in \{+1,-1\}$, where $L$ is the linear system size and $N=4L^3$ the number of sites in the lattice. We then want to calculate the lowest-order energy correction that distinguishes different ground states, that is different configurations of $\{\eta_i\}$. As we discuss in detail in \appref{app:rspt},
such corrections arise from linked clusters of nontrivial topology on the pyrochlore lattice the smallest of which are hexagons [Fig. \ref{fig:ising} (a)]. The leading correction distinguishing configurations of $\{\eta_i\}$ thus occurs at sixth order in RSPT.

Any hexagon on the pyrochlore lattice connects exactly two $\beta$ chains, but there are two inequivalent kinds of hexagon, as shown in \autoref{fig:ising} (a). Neglecting constant terms, that is terms that do not distinguish different chain configurations $\{\eta_i\}$, the energy correction from a hexagon is proportional to the product of the Ising variables of the two chains $i,j$ it connects
\begin{subequations}
\begin{align}
    \delta E^{(6)}_{ij} &= \mathrm{const} + J_{\mathrm{ch}} \eta_i\eta_j, \\
    \delta E^{\prime(6)}_{ij} &= \mathrm{const} + \frac{J_{\mathrm{ch}}'}{2} \eta_i\eta_j,
\end{align}
\end{subequations}
where we use $J_{\mathrm{ch}}$ and $J_{\mathrm{ch}}'$ to distinguish the two ways in which two nearest-neighbor chains can be connected by a hexagon. The factor of $1/2$ here is needed since there are $2L$ equivalent hexagons connecting to chains $i$ and $j$ in the primed direction (indicated in pink in \autoref{fig:geom}). The interaction energy between two $\beta$ chains $i,j$ is then
\begin{equation}
    E^{\mathrm{(int)}}_{ij} = L J_{\mathrm{ch}}^{(\prime)}\,\eta_i \eta_j,
\end{equation}
where $2L$ is the number of sites on the chain.

The $\beta$ chains embedded in the pyrochlore lattice thus form an anisotropic triangular lattice Ising model which can be described using an effective Hamiltonian
\begin{equation}
    \mathcal H_{\mathrm{eff}} = J_{\mathrm{ch}} \sum_{\expval{ij}} \eta_i \eta_j + J_{\mathrm{ch}}' \sum_{\expval{ij}'} \eta_i \eta_j,
    \label{eq:nn-ising-ham}
\end{equation}
where the sum over $\expval{ij}$ runs over nearest neighbor pairs in the [112] and $[\bar1\bar12]$ directions and the sum over $\expval{ij}'$ indicates summation over nearest-neighbor pairs in the [110] direction. Note that here, we use ``nearest-neighbor'' to mean ``connected by a hexagon'' instead of referring to real-space distance since the $\beta$ chains are farther apart in the [110] direction than in the other two directions. We indicate which chain coupling is generated by which linked cluster by matching colors in 
\autoref{fig:ising} (a) and (b). 

The effective couplings $J_{\rm ch}^{(\prime)}$ are functions of the original pyrochlore exchange couplings $J_\lambda$, the field $h$ and the mixing angle $\theta$. The explicit functional form is different in each of the different chain phases. We thus denote the chain couplings in the ${\rm Chain}_\lambda$ phase by $J_{\rm ch}^{\lambda(\prime)}$:
\begin{equation}
    J_{\rm ch}^{(\prime)} = 
        J_{\rm ch}^{\lambda(\prime)}~~\text{in ${\rm Chain}_\lambda$ phase}.
\end{equation}
We calculate the $J_{\rm ch}^{\lambda(\prime)}$ in sixth order RSPT (see \appref{app:rspt} for details on the perturbative expansion). First consider the result in the ${\rm Chain}_Z$ phase
\begin{subequations}
\begin{align}
    J_{\mathrm{ch}}^z =& - S J_z \cos(\phi)^2 \frac{J_x^2 J_y^2}{J_z^4} \frac{\left( J_x - J_y\right)^2}{J_z^2} \frac{1}{16}  \Gamma_z(h),\\
    J_{\mathrm{ch}}^{z\prime} =& S J_z \cos(\phi)^4\frac{J_x^2 J_y^3}{J_z^5} \left( J_x/J_z + \tan(\phi)^2 \right) \frac{2}{16}  \Gamma_z'(h),
\end{align}
\label{eq:effJ-rspt-z}
\end{subequations}
where $S$ is the spin length and the $J_\lambda$ are the exchange couplings in \autoref{eq:hamiltonian}. The angle $\phi$ parameterizes the classical configuration of spins on the $\alpha$ chains such that $\theta-\phi$ is the angle between the local field and the magnetic moments. $\phi$ thus depends on the external field $h, \theta$ as well as on the couplings $J_x,J_z$ and $\phi\to\theta$ as $h \to \infty$. $\Gamma_z$ and $\Gamma_z'$ are dimensionless, positive factors controlling the asymptotic field dependence of the couplings. While full expressions are given in \appref{app:rspt}, we note that the field dependence of $\Gamma_z$ and $\Gamma_z'$ is qualitatively different:
\begin{subequations}
\begin{align}
    \Gamma_z(h)\sim h^{-2} \text{ as } h\to\infty,\\
    \Gamma_z'(h)\sim h^{-4} \text{ as } h\to\infty,
\end{align}
\label{eq:gamma}
\end{subequations}
where the asymptotic exponents corresponds to the number of polarized sites on the hexagon generating the respective coupling [\autoref{fig:geom} (a)]. This enables tuning the order-by-disorder strength and even the ground state by means of the magnetic field (see \autoref{sec:field} for details). 

The respective expressions for $J_{\rm ch}^{x}$ and $J_{\rm ch}^{x\prime}$ are obtained from \autoref{eq:effJ-rspt-z} by interchanging $J_x \leftrightarrow J_y$ and $\cos\phi \leftrightarrow \sin\phi$. It is also given explicitly in \appref{app:rspt}.

In the ${\rm Chain}_Y$ phase, the couplings in sixth order RSPT are given by
\begin{subequations}
\begin{align}
    J_{\rm ch}^y &= -\frac{1}{64} S J_y \sin(2\phi)^2 \frac{J_x^2 J_z^2 (J_x - J_z)^2}{J_y^6} \Gamma_y(h), \\
    J_{\rm ch}^{y\prime} &= 0,
\end{align}
\label{eq:effJ-rspt-y}
\end{subequations}
where $\Gamma_y$ again is a dimensionless, positive, field dependent factor that vanishes as $h^{-2}$ as $h \to \infty$. For $S=1/2$, a finite contribution to $J_{\rm ch}^{y\prime}$ even at higher orders is only generated when including a transverse term corresponding to cubic magnon terms in LSWT.

The phase diagram of the effective model [\autoref{eq:nn-ising-ham}] in terms of the effective couplings $J_{\mathrm{ch}}$ and $J_{\mathrm{ch}}'$ \cite{Dublenych2013} is shown in \autoref{fig:ising} (c). 
There are three possible phases: if $J_{\rm ch}$ is negative and $-J_{\rm ch}>J_{\rm ch}'$ the system orders ferromagnetically. If the exchange parameter $J_{\rm ch}$ is positive and larger than $J_{\rm ch}'$, the system enters a `staggered' phase where there is ferromagnetic order along the primed direction (indicated in \autoref{fig:ising} (b) in pink) with antiferromagnetic order in the perpendicular direction. Finally, if $J_{\rm ch}' > \abs{J_{\rm ch}}$, the system is frustrated since when the system is ordered antiferromagnetically in the primed direction (indicated in \autoref{fig:ising} (b) in pink), it is not possible to satisfy the non-primed bonds (indicated in \autoref{fig:ising} (b) in green). Thus, in this `zigzag' phase, the ground state is highly degenerate, with each antiferromagnetically ordered row retaining an effective Ising degree of freedom. 
Following the ferromagnetic order in the non-primed direction yields the eponymous zigzag shape, indicated in red in the sketch in \autoref{fig:ising} (c).

Since it is clear from \autoref{eq:effJ-rspt-z} and \autoref{eq:effJ-rspt-y} that $J_{\mathrm{ch}}$ is negative for all parameters, the chains will never order in a staggered configuration. This already yields a ferromagnetic ground state across the whole ${\rm Chain}_Y$ phase since there $J_{\mathrm{ch}}'=0$ for $S=1/2$ at sixth order.
In contrast, in the ${\rm Chain}_X$ and ${\rm Chain}_Z$ phases $J_{\mathrm{ch}}'$ is finite and can take either sign. The system is thus either in the ferromagnetic or zigzag phase, depending on sign and strength of $J_{\mathrm{ch}}'$ relative to $J_{\mathrm{ch}}$.
Focusing on the ${\rm Chain}_Z$ phase for simplicity, in terms of the original pyrochlore exchange couplings ($J_\lambda$), if $J_y$ and $J_x + J_z \tan(\phi)^2$ have different signs, $J_{\mathrm{ch}}'$ is negative and we expect the chains to order ferromagnetically.
In contrast, for $J_x = J_y$, $J_{\mathrm{ch}}$ vanishes while $J_{\mathrm{ch}}'$ is positive and finite, at least for $J_x >0$ and $J_x < -J_z \tan^2\phi$ so we expect the chains to order in a zigzag configuration. 
As indicated in \autoref{fig:ising} (c), the phase boundary between the ferromagnetic and zigzag phases is obtained by equating $J_{\mathrm{ch}}' = -J_{\mathrm{ch}}$. This phase boundary as a function of the $J_\lambda$ is also indicated in \autoref{fig:obd-phases-1} and \ref{fig:obd-phases-2} as a solid line.

Note that, while the leading order contribution to the effective chain couplings vanishes in certain regions of the phase diagram, we expect that higher-order terms will contribute also in those regions and eventually lift the classical degeneracy. However, since the degeneracy lifting in these cases is of higher order, it is strongly suppressed.

\begin{figure}
    \centering
    \includegraphics{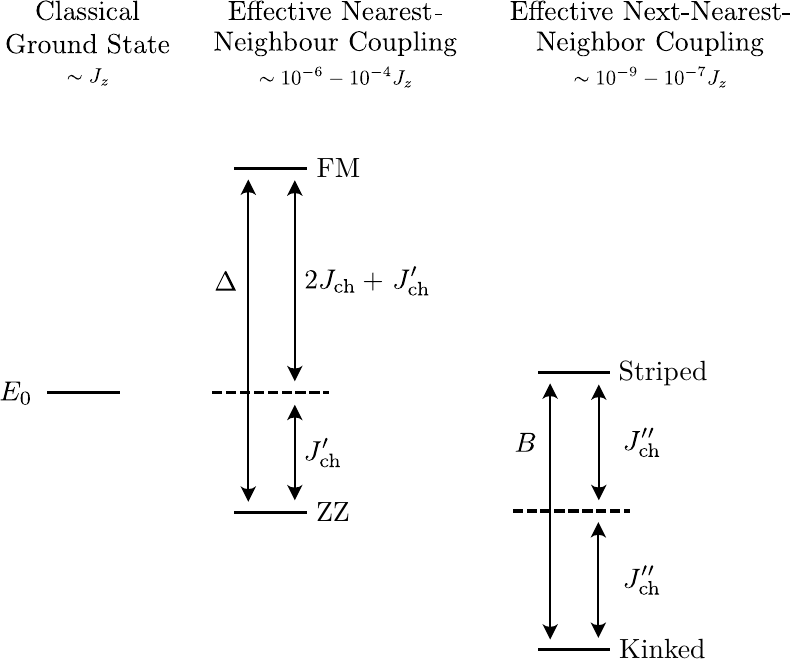}
    \caption{Hierarchy of energy scales in ground state selection in the ${\rm Chain}_Z$ phase. The classical ground state energy is separated from the excited states on a scale set by $J_z$. On an energy scale of $\Delta$, which for most values of exchange couplings lies in a range of $10^{-6} - 10^{-4}J_z$ when $h\sim J_z$, quantum fluctuations split the classical ground state manifold into the ferromagnetic state and the band of zigzag states. The choice of a single zigzag state happens again on a much lower energy scale $B$, which for most values of parameters lies in a range of $10^{-9}-10^{-7}J_z$. The numerical values for the scales are obtained from linear spin wave theory [\autoref{fig:obd-phases-1} and \ref{fig:obd-phases-2}].}
    \label{fig:scales}
\end{figure}

\subsection{Degeneracy lifting in the zigzag phase}

Dipolar-octupolar pyrochlores in a magnetic field undergo a remarkable dimensional evolution when considering different energy scales. On the largest energy scale, corresponding to the bare parameters of the Hamiltonian, the system appears fully three dimensional. In an intermediate regime (between $J_{\rm ch}, J_{\rm ch}'$ and  $J_z$) the system undergoes a dimensional reduction such that it can be described as an ensemble of noninteracting one-dimensional chains. 
At the lowest energy scales, one might expect the order by disorder mechanism to restore the three dimensional nature of the system.

The zigzag phase, however, in some region of the phase diagram adds an additional stage to this dimensional evolution. Viewed in the pyrochlore lattice, left-over degrees of freedom are effectively two-dimensional. Magnetic moments on the $\beta$ chains are ordered in stripe patters within the $[hl0]$ plane, however the effective interactions between the planes are frustrated. Thus, the system has a ground state with a subextensive degeneracy scaling with the linear system size $L$, each plane carrying an effective Ising degree of freedom.

This degeneracy is broken only at 10th order in perturbation theory, by a linked cluster indicated by the blue dashed line in \autoref{fig:ising} (a). This cluster generates a next-nearest-neighbor coupling $J_{\rm ch}''$ in the $[001]$ direction as indicated in \autoref{fig:ising} (b) and, equivalently, a ring exchange term. Note that the other possible next-nearest-neighbor coupling, while also generated at 10th order, is still frustrated in the zigzag phase and does not lift the degeneracy. As shown in \autoref{fig:ising} (d), depending on the sign of the next-nearest-neighbor coupling $J_{\mathrm{ch}}''$, there are two different zigzag states selected, either a `kinked' state, where ordered moments form a zigzag pattern in the [001] direction, or a `striped' state, where magnetic moments order ferromagnetically along one of the non-primed nearest-neighbor directions.

The `ferromagnetic', `kinked' and `striped' configurations all have only an intensive number of degenerate states related by symmetries of the Hamiltonian [\autoref{eq:hamiltonian}]. Hence, the system in the zigzag phase regains its full three dimensional correlations on an energy scale $\sim J_{\rm ch}''$ while on scales between $J_{\rm ch}''$ and $J_{\rm ch}, J_{\rm ch}'$ it features effectively two-dimensional degrees of freedom.

\begin{figure*}
    \centering
    \includegraphics{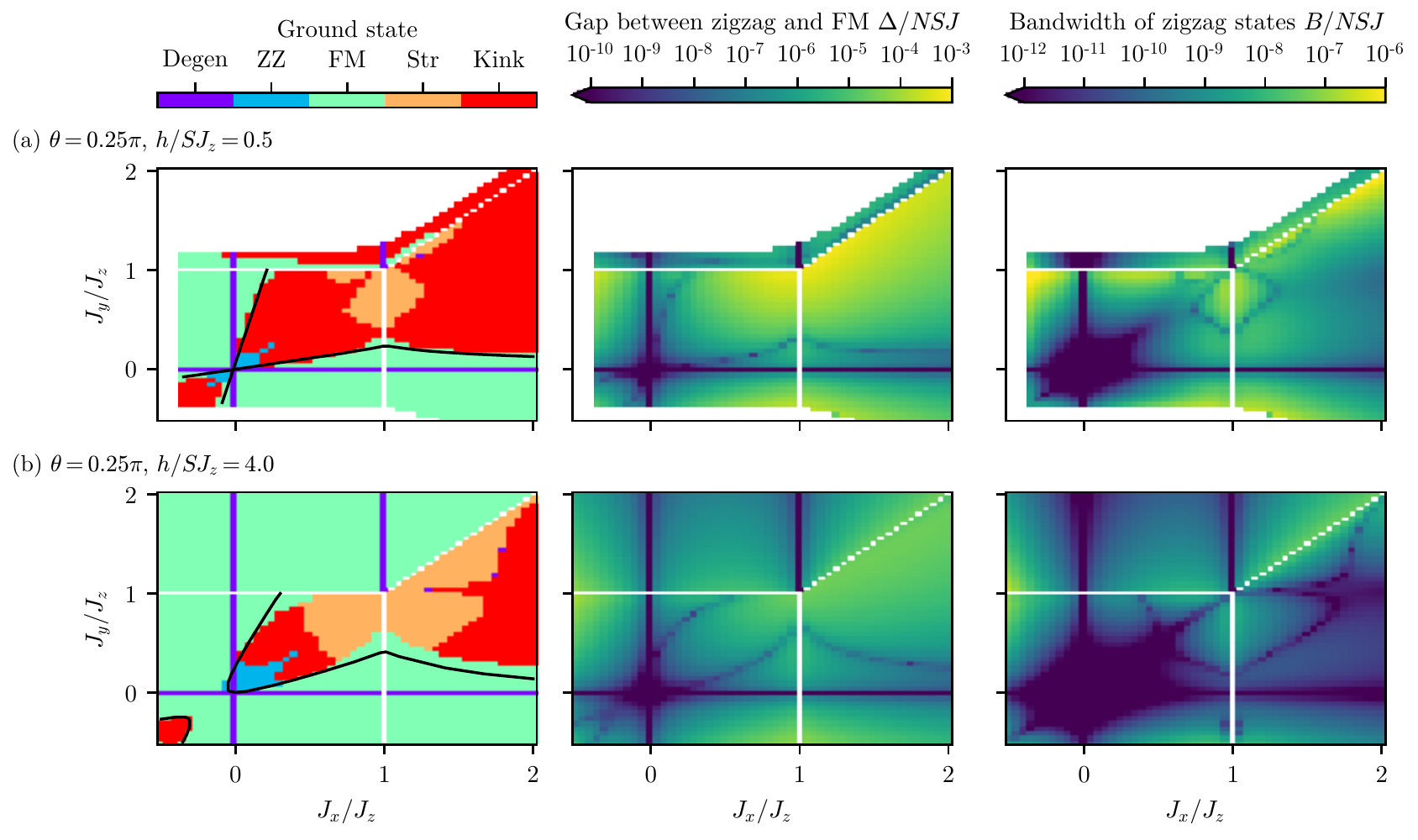}
    \caption{Order-by-disorder (OBD) phase diagram of dipolar-octupolar pyrochlores for mixing angle $\theta=\pi/4$, together with the two relevant energy scales of the ground state selection, as a function of exchange parameters. We indicate the ground state found by linear spin wave theory (LSWT) by color. Then, `Degen' denotes that all ground states are degenerate, i.e. no OBD is observed. `ZZ' denotes that a zigzag state is selected, but that the zigzag states are degenerate within the numerical precision (cf. the right column in the respective regions). `FM' denotes that a ferromagnetic chain configuration is selected. The striped (`Str') and kinked (`Kink') configuration are both zigzag configurations
    [\autoref{fig:ising}(d)]. We also indicate the transition between the FM and ZZ phases as computed from real space perturbation theory (RSPT) by a black line.
    We plot the phase diagram for two field values $h=SJ_z/2$ (a) and $h=4SJ_z$ (b). As anticipated from real space perturbation theory, the extent of the zigzag phase is significantly reduced as the external field is increased. This makes it possible to tune between different ground states in some regions of the phase diagram. In all cases, it is evident that the energy scales of ground state selection between the ferromagnetic and the zigzag states are much larger than those of ground state selection within the set of zigzag states.}
    \label{fig:obd-phases-1}
\end{figure*}

\begin{figure*}
    \centering
    \includegraphics{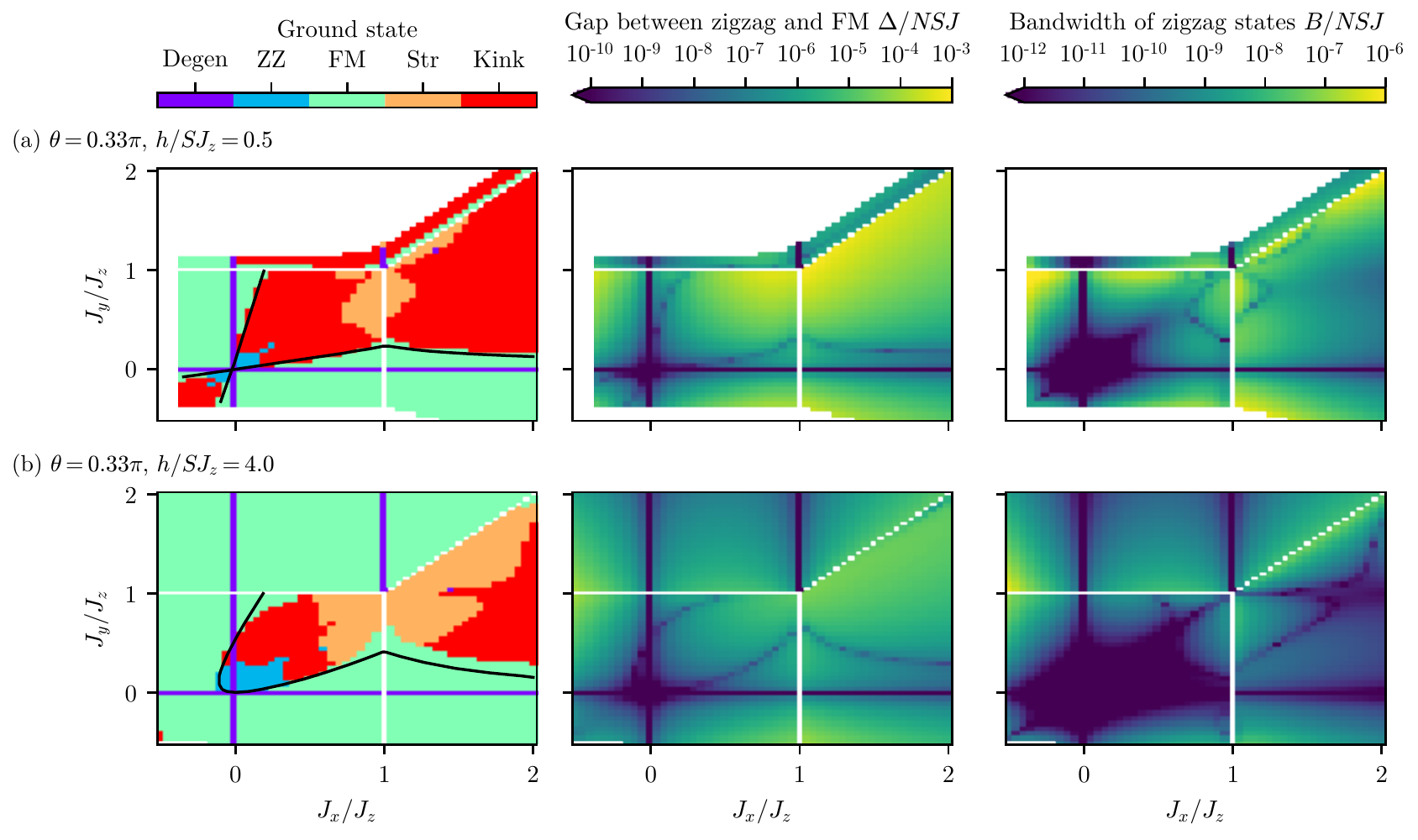}
    \caption{Order-by-disorder (OBD) phase diagram of dipolar-octupolar pyrochlores for mixing angle $\theta=0.33\pi$, together with the two relevant energy scales of the ground state selection, as a function of exchange parameters. We indicate the ground state found by linear spin wave theory (LSWT) by color. Then, `Degen' denotes that all ground states are degenerate, i.e. no OBD is observed. `ZZ' denotes that a zigzag state is selected, but that the zigzag states are degenerate within the numerical precision (cf. the right column in the respective regions). `FM' denotes that a ferromagnetic chain configuration is selected. The striped (`Str') and kinked (`Kink') configuration are both zigzag configurations
    [\autoref{fig:ising}(d)]. We also indicate the transition between the FM and ZZ phases as computed from real space perturbation theory (RSPT) by a black line.
    We plot the phase diagram for two field values $h=SJ_z/2$ (a) and $h=4SJ_z$ (b). As anticipated from real space perturbation theory, the extent of the zigzag phase is significantly reduced as the external field is increased. This makes it possible to tune between different ground states in some regions of the phase diagram. In all cases, it is evident that the energy scales of ground state selection between the ferromagnetic and the zigzag states are much larger than those of ground state selection within the set of zigzag states.}
    \label{fig:obd-phases-2}
\end{figure*}

We summarize the resulting hierarchy of energy scales in ground state selection in \autoref{fig:scales}, where we also give rough numerical ranges in which the scales $\Delta$ and $B$ lie for most values of exchange couplings if $h \sim J_z$ [cf. \autoref{fig:obd-phases-1} and \ref{fig:obd-phases-2}].

\section{Order by Disorder Phase Diagram\label{sec:obd-phases}}

The classical phase diagram of dipolar-octupolar pyrochlores in a moderate external field in the [110] direction, as discussed in \autoref{sec:classical-phases}, contains extended regions of chain phases in which the field generates effective one-dimensional degrees of freedom and which feature a subextensive degeneracy. Since the degenerate states are not related by symmetries, one expects the degeneracy to be lifted by quantum fluctuations. 
In most of these chain phases (the ${\rm Chain}_\lambda$ phases), this degeneracy lifting, as established in \autoref{sec:eff-Ising}, can be cast in the form of an anisotropic triangular-lattice Ising model with three effective chain coupling parameters. Motivated from the effective model, which was derived in leading-order real space perturbation theory (RSPT), we expect that on the largest energy scale, two different types of chain configuration will be chosen as a ground state. The chains will order either ferromagnetically, or in a zigzag state [\autoref{fig:ising} (c)]. In the latter case, there is a further dimensional evolution to effectively two-dimensional degrees of freedom, which are planes perpendicular to the [001] direction, each carrying an effective Ising degree of freedom, noninteracting in leading-order RSPT. An effective interaction $J_{\rm ch}''$, breaking the residual degeneracy, is only generated at 10th order
of RSPT. Depending on the sign of $J_{\rm ch}''$, one of = two zigzag configurations will be chosen, either a `kinked' or a `striped' state both of which then feature an intensive number of degenerate states related by symmetries [\autoref{fig:ising} (d)].

\subsection{Phase diagram}

In this section, we turn to calculate the full classical degeneracy lifting in the ${\rm Chain}_\lambda$ phases, as a function of the exchange parameters $J_\lambda$. To corroborate the RSPT calculation, we calculate the lifting numerically using linear spin wave theory (LSWT). The calculation is performed using a semi-infinite slice of a pyrochlore lattice, that is a $8\times 8$ triangular lattice of infinite $\beta$ chains with periodic boundary conditions in the directions orthogonal to the chains.

The central quantity to compute ground state selection by quantum fluctuations is the zero point energy
\begin{align}
    \mathcal E_0 \left( \{\vec S_i^{(0}\} \right) = \frac{1}{\pi} \sum_\nu \int_0^{\pi/2} \dd k\, \omega_\nu \left((k, -k, 0)\right),
\end{align}
where $\omega_\nu(\vec q)$ are the magnon frequencies of the system around the ground state $\{\vec S_i^{(0)}\}$. Details of the numerical computation can be found in \appref{app:lswt}.
We compute the zero point energy for all possible ground state configurations of the effective nearest-neighbor model, which are the ferrromagnetic state, the staggered state and 12 different zigzag states (those are all zigzag states on a $8\times 8$ triangular lattice not related by lattice symmetries). 
Assuming that the effective model is a valid low-energy description of the system, this will yield the correct ground state phase diagram and serve as a quantitative test of the effective couplings calculated in RSPT. 
To test the validity of the effective model, we also compute the zero point energies for $1000$ randomly selected chain configurations and compare them to the predictions of the effective model, with three couplings fitted using LSWT (see \autoref{sec:obd-phases-b} for details). We find that the effective model indeed gives an accurate estimate of the energies, and therefore that searching for ground states only among the possible ground states of the effective model is justified. 

The resulting (quantum) order-by-disorder phase diagram for the ${\rm Chain}_\lambda$ ($\lambda = x,y,z$) phases is shown in \autoref{fig:obd-phases-1} and \autoref{fig:obd-phases-2} for $\theta=0.25\pi$ and $\theta=0.33\pi$ respectively. The selected configurations are those expected from the effective Ising model. Remarkably, the phase boundaries obtained from LSWT also match very well with those obtained directly from RSPT at least for moderate transverse couplings. 
Going beyond the RSPT calculation, LSWT also reveals the splitting of the zigzag state into the striped and kinked configurations.
There are also parts of the phase diagram where the classical degeneracy is not lifted within LSWT. Most notably, this occurs along the two lines $J_x = 0$ and $J_y = 0$ for which we prove explicitly that $\omega(\vec q)$ are identical for all classical ground states within LSWT in \appref{app:lswt}.

To shed light on the relevant energy scales in the ground state selection by quantum fluctuations, we also show in \autoref{fig:obd-phases-1} and \ref{fig:obd-phases-2} the gap between the ferromagnetic and the zigzag configurations $\Delta$ as well as the bandwidth of the zigzag states $B$ (that is also the gap between the striped and the kinked state).
It is evident that the energy scales on which the classical degeneracy is lifted are very small and have a clear hierarchy as illustrated in \autoref{fig:scales}. 
In particular, for the compound Nd$_2$Zr$_2$O$_7$, using the most recent estimate of the exchange couplings\cite{Xu2019} and a field of about 1 Tesla yields $\Delta \approx 10^{-8}\,\mathrm{meV}$. In this example, quantum OBD has thus no experimental relevance in ground state selection. 

To investigate the influence of finite temperature on OBD, we employ two approaches (see \autoref{app:temperature} for details). First, we compute the leading order contribution of the classical low-temperature expansion of the free energy and show that it is identical for all classical ground states in the Chain$_\lambda$ phases. Second, we compute the free energy of the noninteracting magnon gas at finite temperature and show that below a crossover temperature $T_{\rm co}$ set by the spin wave gap, the energy scales $\Delta$ and $B$ are barely modified from their zero temperature values. For Nd$_2$Zr$_2$O$_7$, we estimate $T_{\rm co} \approx 0.23 K$ (see \autoref{fig:obt-tempterature}).

Our findings are consistent with the fact that disordered chain configurations were observed in neutron scattering experiments on this compound \cite{Xu2018} at low temperatures.

\subsection{Validity of effective model\label{sec:obd-phases-b}}

In order to further assess the validity of the effective model, we compare it to the full spectrum of zero point energies.
To this end, we compute the effective chain couplings $J_{\rm ch}$ and $J_{\rm ch}'$ explicitly using the zero point energies obtained from LSWT using
\begin{align}
    J_{\rm ch} &= \frac{1}{4}\left( \mathcal E_0^{\rm FM} - \mathcal E_0^{\rm Staggered}\right), \nonumber \\
    J_{\rm ch}' &= \frac{1}{2}\left( \mathcal E_0^{\rm FM} + \mathcal E_0^{\rm Staggered}\right), \nonumber \\
    J_{\rm ch}'' &= \frac{1}{2}\left( \mathcal E_0^{\rm Kinked} - \mathcal E_0^{\rm Striped}\right).
    \label{eq:eff-j-lswt}
\end{align}
We then compare the zero point energies $\mathcal E_0(\{\vec S_i^{(0)} \})$ of all $1014$ classical ground state configurations that we consider (ferromagnet, staggered, 12 zigzag states and 1000 randomly chosen configurations) to the estimate from the effective nearest-neighbor model using the effective couplings obtained from Eq. \ref{eq:eff-j-lswt}. 

Within the ${\rm Chain}_Z$ and ${\rm Chain}_X$ phases, the estimates agree remarkably well with the energies computed from LSWT, with residues between the two on the order of the next nearest-neighbor coupling $J_{\rm ch}''$ [see \autoref{fig:eff-model-res} in \appref{app:lswt}].
Restricting ourselves to the 12 zigzag states, their splitting is modeled by the fitted $J_{\rm ch}''$ up to $10^{-12} J_z$.

In the ${\rm Chain}_Y$ phase, we expect the effective model to be still valid as derived in RSPT. At high fields, both LSWT and RSPT predict dominance of the ferromagnetic configuration for all exchange couplings in that phase. However, at low fields LSWT for some part of the phase diagram predicts a zigzag ground state, in contrast with RSPT. 
The disagreement can be rationalized by considering that LSWT is unable to capture the hard-core constraint of magnons for $S=1/2$ and this constraints leads the coupling $J_{\rm ch}'$ to vanish at sixth order in RSPT.
Meanwhile, in RSPT we neglect higher orders which could lead to a finite $J_{\rm ch}'$ even for $S=1/2$.

\section{Field tunability\label{sec:field}}

\begin{figure}
    \centering
    \includegraphics{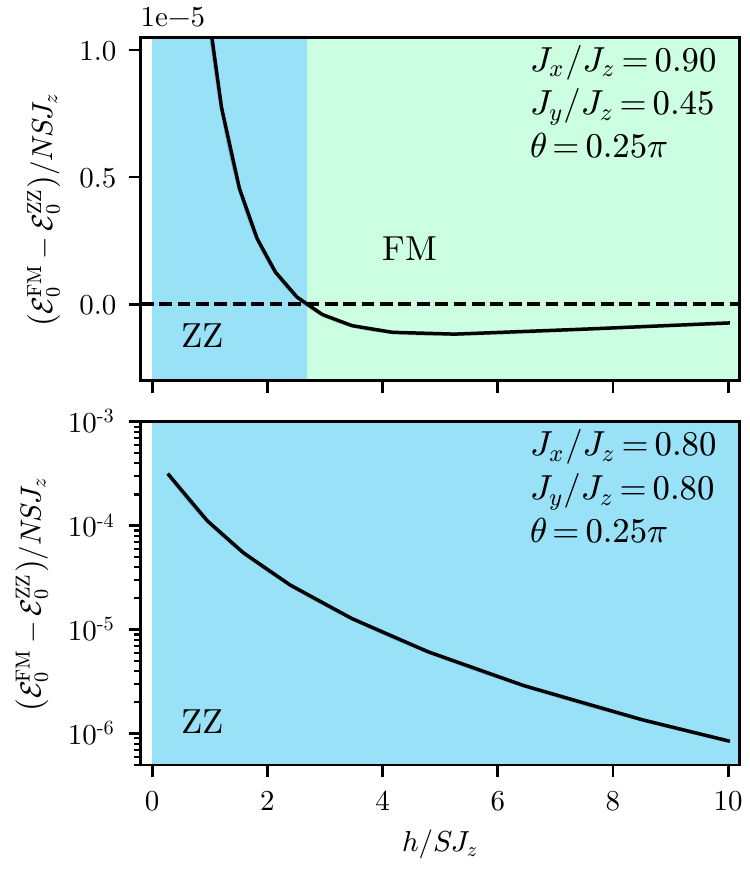}
    \caption{Difference of the zero point energy of the ferromagnetic chain configuration and the zigzag configuration from LSWT as a function of field for fixed exchange parameters. Top: for $J_x = 0.9 J_z$, $J_y = 0.45 J_z$, it is possible to tune the ground state of the system by means of the external field. Bottom: for $J_x = J_y = 0.8 J_z$, the ground state is always a zigzag configuration, however the OBD energy scale is much larger. 
    The splitting of the zigzag states in both cases is imperceptible on the scale of the plot.}
    \label{fig:zpe-h}
\end{figure}

As is qualitatively clear from \autoref{eq:gamma}, the effective nearest-neighbor couplings depend on the external magnetic field $h$. 
Since the effective interaction between chains is generated by fluctuations on the $\alpha$ chains, the external field can directly control the strength of these fluctuations and hence be used to manipulate the OBD effect. This is clearly seen in \autoref{fig:zpe-h}, where the difference of the zero point energies of the ferromagnetic state and the zigzag states is shown as a function of field for two sets of exchange parameters.

Furthermore, the ground state that is selected by quantum fluctuations can itself be tuned by means of the field. This is the case because selection of the ferromagnetic state or the zigzag states is determined by a competition of the effective couplings in the two inequivalent nearest-neighbor directions of the lattice. Inspecting the two leading-order contributions shown as green and pink hexagons in \autoref{fig:ising} (a), we see that the two hexagons have two and four sites located on $\alpha$ chains respectively. Hence, as is reflected in the different behaviour of $\Gamma_z$ and $\Gamma_z'$ in \autoref{eq:gamma}, the couplings scale differently with the field and hence the competition between $J_{\rm ch}$ and $J_{\rm ch}'$ changes as a function of field. As an example of this, we show in the top panel of \autoref{fig:zpe-h} how the energy difference between the ferromagnetic state and the zizag states (computed using LSWT) changes sign as a function of field.

\section{Experimental relevance\label{sec:experiment}}

In this Section we discuss the implications of our results for the dipolar-octupolar pyrochlore compounds Nd$_2$Zr$_2$O$_7$ and Ce$_2$Sn$_2$O$_7$.

\subsection{Disordered dipolar chain state in \texorpdfstring{Nd$_2$Zr$_2$O$_7$}{Nd2Zr2O7}}

First, a transition from an ${\rm AIAO}$ ordered phase to a disordered ${\rm chain}$ phase induced by a field has been observed in Nd$_2$Zr$_2$O$_7$ \cite{Xu2018}. Such a transition is expected for $-J_x < J_z < -J_x/3$ or $-J_z < J_x < -J_z/3$, which is consistent with the exchange parameters of this compound as estimated from neutron scattering, for which the most recent estimate (taken from Ref. \onlinecite{Xu2019}) is 
\begin{align}
    J_x \approx 0.1\,\mathrm{meV}, ~ J_z \approx -J_x/2, ~ J_y \approx 0.15 J_x.
    \label{eq:params-Nd2Zr2O7}
\end{align}

For these exchange parameters and assuming a field of 1~Tesla, LSWT predicts a ferromagnetic order of the chains that is however separated from the zigzag state by a gap of only $\Delta \approx 10^{-8}\,\mathrm{meV}$. At temperatures above $T_{\rm co}\approx 0.23\,K$, this is even further suppressed (see discussion in \autoref{app:temperature}). Hence, even at the lowest experimentally accessible temperatures one would not expect to see ordering of the chains driven by quantum fluctuations, which is consistent with the experimental observation of a disordered chain state.

This makes Nd$_2$Zr$_2$O$_7$ in a moderate [110] field an excellent platform for the study of one-dimensional quantum XYZ chains.

\subsection{Octupolar quantum spin liquid in \texorpdfstring{Ce$_2$Sn$_2$O$_7$}{Ce2Sn2O7}}

\begin{figure}
    \centering
    \includegraphics{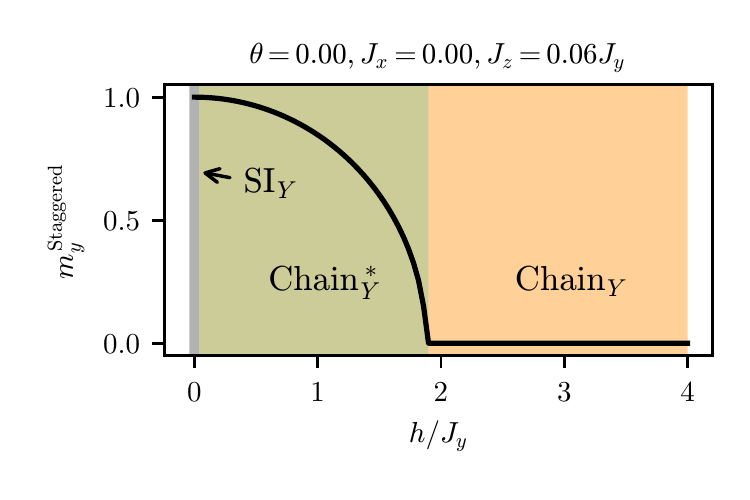}
    \caption{Transition from the ${\rm Chain}_Y^*$ to the ${\rm Chain}_Y$ phase as a function of the external field, for exchange parameters used to model Ce$_2$Sn$_2$O$_7$ in Ref. \onlinecite{Sibille2020}. The plot shows the staggered octupolar moment on the $\alpha$ chains [\autoref{eq:staggered-m}], which is nonzero only in the ${\rm Chain}_Y^*$ phase. Classically, the octupolar ice phase ${\rm SI}_Y$ exists only for strictly zero field $h=0$, however in the quantum case there would be a finite
    window at low field with a quantum spin liquid ground state.}
    \label{fig:chain-y-h}
\end{figure}

Second, recent experimental evidence for an octupolar $U(1)$ quantum spin liquid (QSL) has been reported in the dipolar-octupolar pyrochlore Ce$_2$Sn$_2$O$_7$ \cite{Sibille2020}. This implies that in this compound the octupolar exchange is positive and large \cite{Li2017, Patri2020, Yao2020, Benton2020}
\begin{align}
    J_y > \max(J_x, J_z, -3J_x, -3J_z).
\end{align}

Classically, this would result in an ice-like phase of octupolar moments ${\rm SI}_Y$ at zero-field. This phase has the same extensive degeneracy as spin-ice. At finite fields, the system would be in the ${\rm Chain}_Y^*$ phase in which the extensive degeneracy is broken down to a subextensive degeneracy with one Ising degree of freedom carried by each $\alpha$ and each $\beta$ chain [\autoref{fig:geom}]. At a finite critical field, the system would then undergo a second-order transition to the ${\rm Chain}_Y$ phase, which also features a subextensive degeneracy but with only half as many Ising degrees of freedom, located on the $\beta$ chains. 
This is shown in \autoref{fig:chain-y-h}, where we plot the (classical) octupolar staggered moment [\autoref{eq:staggered-m}] as a function of the external field for exchange parameters as used to model Ce$_2$Sn$_2$O$_7$ in Ref. \onlinecite{Sibille2020}.
The second order transition at finite field is absent if the ground state is a dipolar spin ice state ${\rm SI}_X$ and ${\rm SI}_Z$ since those can be directly connected to the respective chain phases ${\rm Chain}_X$ and ${\rm Chain}_Z$.

In a real compound, where $S=1/2$, quantum fluctuations might be expected to be strong, the story laid out above gets slightly modified. The zero field octupolar ice phase ${\rm SI}_Y$ will be replaced by a $U(1)$ QSL that prevails for some finite range of external field\cite{Hermele2004}. At some critical field $h_c^{(1)}$ this is followed by a confinement transition to a quantum version of the ${\rm Chain}_Y^*$ phase. At a larger, second critical field $h_c^{(2)}$, there will then be a transition to the `simple' chain phase ${\rm Chain}_Y$. 

The transition from the QSL to ${\rm Chain}_Y^*$ can be captured  by means of degenerate perturbation theory around the low field limit. As we show in \appref{app:quantum-chain-y-star}, for $S=1/2$ there is a diagonal perturbation at fourth order in the external field that generates an effective antiferromagnetic (with respect to the local basis) interaction within the $\alpha$ chains. 
This interaction favours a set of states with one-dimensional degrees of freedom on each $\alpha$ and $\beta$ chain, corresponding precisely to the ${\rm Chain}_Y^*$ phase found in our classical analysis.
This indicates that the ${\rm Chain}_Y^*$ phase exists in the quantum limit, and should appear in octupolar quantum spin ices subject to moderate [110] fields.

The double transition from a $U(1)$ QSL to the ${\rm Chain}_Y^*$ phase and finally to the ${\rm Chain}_Y$ phase as a function of a [110] field appears to be unique to the case of dominantly octupolar exchange and can hence serve as another experimental test for compounds suspected to realize such physics like Ce$_2$Sn$_2$O$_7$.
Additionally, the existence of the ${\rm Chain}_Y^*$ phase is interesting in and of itself. It serves as an example of how uniform external fields acting on noncollinear magnetic states can lead to the development of complex degrees of freedom. While the number of degrees of freedom grows only with the square of the linear system size $L^2$, their interaction should not be expected to be modeled by a local two-dimensional effective model since $\alpha$ and $\beta$ chains interpenetrate in a nontrivial way.

\section{Conclusion\label{sec:conclusion}}

In conclusion, we have mapped out the classical phase diagram of dipolar-octupolar pyrochlores in a [110] field in detail and studied quantum order-by-disorder (OBD) in the `chain' phases, in which the classical ground state degeneracy can be parametrized by a subextensive number of effective Ising degrees of freedom, each describing the magnetic moment of a one-dimensional chain. We have focused our study on a subset of these phases, where all degrees of freedom are carried by a set of parallel chains. We show that OBD in this case can be modelled remarkably well by a simple effective anisotropic triangular lattice Ising model that we derive in real space perturbation theory (RSPT) and corroborate using numerical linear spin wave theory (LSWT). 

Ground state selection by quantum fluctuations in dipolar-octupolar pyrochlores is a multi-step process governed by a hierarchy of energy scales. First, competition of two inequivalent nearest-neighbor chain interactions drives selection of either a ferromagnetic ordering of the chains or a `zigzag' order. In the latter case, the chains are ordered antiferromagnetically in one direction while the nearest-neighbor coupling in the other two directions is frustrated, leaving a subextensive degeneracy scaling with the linear system size. This left-over degeneracy is finally broken by an effective next-nearest-neighbor coupling, albeit on a minuscule energy scale.
Taken together, the system undergoes a highly nontrivial dimensional evolution when considering effective degrees of freedom on different energy scales.
Comparing the results of the effective model with LSWT, the two methods mostly agree remarkably well even quantitatively, with phase boundaries between ferromagnetic and zigzag phases almost indiscernible in the dipolar chain phases ${\rm Chain}_X$, ${\rm Chain}_Z$ for moderate transverse couplings. 
%Ground state selection can thus be fully captured by an effective model with only three parameters, which in the dipolar chain phases also agree qualitatively when derived using two different methods.

Highlighting connection to different experiments, we have also discussed that the energy scales on which quantum fluctuations break the classical degeneracies are minuscule for parameter values relevant to Nd$_2$Zr$_2$O$_7$. This is consistent with the experimental observation of a disordered chain state in this compound \cite{Xu2018} and makes this material a promising platform to study pure quantum XYZ chains. 
Finally, we study the field-breakdown of the octupolar quantum spin liquid (QSL) state recently observed in Ce$_2$Sn$_2$O$_7$. We predict that for a field in the [110] direction, the transition to the large-field chain state ${\rm Chain}_Y$ happens in two steps, via an intermediate phase that we call ${\rm Chain}_Y^*$. This sequence of transitions is unique for the case of dominantly octupolar exchange and can hence be used as another test for the nature of Ce$_2$Sn$_2$O$_7$.

{Acknowledgements:}
This work was in part supported by the Deutsche Forschungsgemeinschaft  under grants SFB 1143 (project-id 247310070) and the cluster of excellence ct.qmat (EXC 2147, project-id 390858490).

\appendix

\section{Derivation of leading-order correction in real space perturbation theory\label{app:rspt}}

\subsection{Setup}

In this appendix, we derive the expression for the leading-order effective chain couplings presented in \autoref{sec:eff-Ising} using real space perturbation theory. We start from a Hamiltonian of the form
\begin{equation}
    \mathcal H = \sum_{\expval{ij}} \vec S_i \cdot \vec J \cdot \vec S_j - \sum_i \vec h_i \cdot \vec S_i,
    \label{eq:app1-H0}
\end{equation}
from which \autoref{eq:hamiltonian} is obtained for the choice $\vec J = \mathrm{diag}(J_x, J_y, J_z)$ and $\vec h_i = \mu_{\mathrm{B}} g_z \vec H \cdot \unitvec z_i \left(\sin\theta, \ 0, \ \cos\theta \right)$,
with ${\bf H} \parallel (1,1,0)$. 

We transform to a local basis, $\unitvec u_i$, $\unitvec v_i$, $\unitvec w_i$ chosen such that $\vec S_i^{(0)} = S\unitvec w_i$ is a classical ground state.
The Hamiltonian is then separated an unperturbed part, $\mathcal{H}_0$ and four transverse perturbations
\begin{equation}
    \mathcal H = \mathcal H_0 + \sum_{\expval{ij}} V_{ij}^{(1)} + V_{ij}^{(2)} + V_{ij}^{(3)} +V_{ij}^{(4)},
\end{equation}
where
\begin{subequations}
\begin{flalign}
    \mathcal H_0 &= E_0 + \sum_i B_i \delta S_i, \\
    E_0 &= \frac{S^2}{2} \sum_i \left( \sum_{j\in \mathrm{nn}(i)} \unitvec{w}_i \cdot \vec J \cdot \unitvec{w}_j \right) - S \unitvec{w}_i \cdot \vec h_i, \\
    \vec B_i &= -S \sum_{j\in \mathrm{nn}(i)} \vec J \cdot \unitvec{w}_j + \vec h_i,
\end{flalign}
\end{subequations}
and
\begin{subequations}
\begin{align}
    V_{ij}^{(1)} &= \vec c_i^* \cdot \vec J \cdot \vec c_j^*\, S^+_i S^+_j + h.c.,
    \\
    V_{ij}^{(2)} &= \vec c_i^* \cdot \vec J \cdot \vec c_j\, S^+_i S^-_j + h.c.,
    \\
    V_{ij}^{(3)} &= \unitvec w_i \cdot \vec J \cdot \unitvec w_j\, \delta S_i \delta S_j,
    \\
    V_i^{(4)} &= \vec c_i \cdot \vec J \cdot \unitvec w_i\, S^-_i \delta S_j + \unitvec w_i \cdot \vec J \cdot \vec c_j\, \delta S_i S^-_j + h.c.,
\end{align}
\label{eq:app1-V}
\end{subequations}
where $\delta S_i = S - S_i^w$ and $\vec c_i = \tfrac{1}{2}\left(\unitvec u_i + i \unitvec v_i\right)$.

Going forward, we will first calculate the effective couplings in the ${\rm Chain}_Z$ phase (see \autoref{sec:classical-phases}), with the corresponding results for the ${\rm Chain}_X$ phase being obtained by the same procedure. 
Finally, the effective couplings in the ${\rm Chain}_Y$ phase will be computed in the last part of this appendix.

For a ground state in the ${\rm Chain}_Z$ phase (see \autoref{sec:classical-phases}) we have
\begin{subequations}
\begin{align}
    \unitvec w_i =
    \begin{cases}
        \left(\sigma_i^{\alpha}\sin\phi, 0, \sigma_i^{\alpha}\cos\phi\right) & \text{$i\in\alpha$ chain}\\
        \left(0, 0, \sigma_i^{\beta}\eta_c \right) & \text{$i\in\beta$ chain}
    \end{cases},
\end{align}
and choose
\begin{align}
    \vec c_i = \frac{1}{2}
    \begin{cases}
        \left(\sigma_i^{\alpha}\cos\phi, i, -\sigma_i^{\alpha}\sin\phi\right) & \text{$i\in\alpha$ chain}\\
        \left(1, i\sigma_i^{\beta}\eta_c, 0 \right) & \text{$i\in\beta$ chain}
    \end{cases},
\end{align}
\label{eq:app1-basis}
\end{subequations}
where $\eta_c$ is the Ising variable of the chain that the site $i$ belongs to
\begin{subequations}
\begin{align}
    \sigma_i^{\alpha} &=
    \begin{cases}
        1 & \text{if } i \in \mathcal L_0 \\
        -1 & \text{if } i \in \mathcal L_3
    \end{cases},\\
    \sigma_i^{\beta} &=
    \begin{cases}
        1 & \text{if } i \in \mathcal L_1 \\
        -1 & \text{if } i \in \mathcal L_2
    \end{cases}.
\end{align}
\label{eq:app1-sigma}
\end{subequations}
The angle $\phi$ above parameterizes the classical configuration of spins on the $\alpha$ chains such that $\theta-\phi$ is the angle between the local field and the magnetic moments. $\phi$ thus depends on the external field $h, \theta$ as well as on the couplings $J_x,J_z$ and $\phi\to\theta$ as $h \to \infty$.

The perturbations $V_{ij}^{(1\dots4)}$ hence depend parametrically on the chain configuration $\{\eta_i\}$ via the basis \autoref{eq:app1-basis}. Standard perturbation theory yields a configuration dependent energy correction $\Delta E\left(\{ \eta_i \}\right)$. As we will see below, the leading order contribution to $\Delta E$ depends only on products of nearest-neighbor chains such that
\begin{equation}
    \Delta E = \sum_{\expval{ij}} \delta E_{ij} =  J_{\rm ch} \sum_{\expval{ij}} \eta_i\eta_j + J_{\rm ch}' \sum_{\expval{ij}'} \eta_i\eta_j,
\end{equation}
which yields the effective Ising model discussed in \autoref{sec:eff-Ising}.

\subsection{Perturbation theory}

Since the energy corrections must stay extensive, any term that contributes to perturbation theory must correspond to a linked cluster of the pyrochlore lattice, where each edge represents one or more applications of the perturbation $V = V^{(1)} + V^{(2)} + V^{(3)} + V^{(4)}$. The energy correction corresponding to a linked cluster with $p$ edges is \cite{Chernyshev2014}
\begin{equation}
    \delta E^{(p)}  = \sum_{\{\psi_i\}} \frac{\bra{0} V \ket{\psi_1} \bra{\psi_1} V \ket{\psi_2} \dots \bra{\psi_{p-1}} V \ket{0}}{(E_0 - E_{\psi_1}) \dots (E_0 - E_{\psi_{p-1}})},
    \label{eq:app1-eval}
\end{equation}
where $E_0$ is the classical ground state energy and $E_{\psi_i}$ is the unperturbed energy of the intermediate state $\ket{\psi_i}$. From the above we can read off some simple rules:
\begin{enumerate}
    \item Any linked cluster corresponds to a sum of terms, which are all possible combinations of applications of the $V_{ij}^{(1\dots4)}$ to the links such that we begin and end in the vacuum of spin flips $\ket{0}$.
    \item Any such sequence of perturbations has to begin and end with either $V^{(1)}$ or $V^{(4)}$.
    \item Any such sequence must involve an even number of applications of $V^{(4)}$.
\end{enumerate}
Since we are interested only in the leading order contribution to the energy correction we do not have to consider the perturbations $V^{(3)}$ and $V^{(4)}$ since the operator $\delta S_i$ does not modify the spin configuration. Thus for any sequence of applications of perturbations involving $V^{(3)}$ or $V^{(4)}$, there will be a nonvanishing lower order term that contributes.

The matrix elements of the perturbations are evaluated straightforwardly using \autoref{eq:app1-V} and \autoref{eq:app1-basis}. If $i,j$ are both on a $\beta$ chain
\begin{subequations}
\begin{align}
    V_{ij}^{(1)} &= \frac{1}{4} \left( J_x + J_y\right) S_i^+ S_j^+ + h.c., \\
    V_{ij}^{(2)} &= \frac{1}{4} \left( J_x - J_y\right) S_i^+ S_j^- + h.c..
\end{align}
In contrast, if $i,j$ are both on an $\alpha$ chain 
\begin{align}
    V_{ij}^{(1)} &= -\frac{1}{4} \left(\cos\phi^2\,J_x + J_y + \sin\phi^2\,J_z\right) S_i^+ S_j^+ + h.c., \\
    V_{ij}^{(2)} &= \frac{1}{4} \left(-\cos\phi^2\,J_x + J_y - \sin\phi^2\,J_z\right) S_i^+ S_j^- + h.c.. 
\end{align}
Finally, if $i$ is on a $\beta$ chain and $j$ is on an $\alpha$ chain
\begin{align}
    V_{ij}^{(1)} &= \frac{1}{4} \left(\sigma^{\alpha}_j \cos\phi J_x - \sigma^{\beta}_i\eta_c J_y\right) S_i^+ S_j^+ + h.c., \\
    V_{ij}^{(2)} &= \frac{1}{4} \left(\sigma^{\alpha}_j \cos\phi J_x + \sigma^{\beta}_i\eta_c J_y\right) S_i^+ S_j^- + h.c..
\end{align}
\label{eq:app1-matrix-elem-Vij}
\end{subequations}

Similarly, the factors in the denominator of \autoref{eq:app1-eval} are given by
\begin{align}
    E_0 - E_{\psi} &= \bra{\psi}E_0 - H_0\ket{\psi} \nonumber\\
        &= -\sum_i B_i \bra{\psi}\delta S_i \ket{\psi},
\end{align}
where if $i$ is on a $\beta$ chain:
\begin{subequations}
\begin{align}
    B_i = 2 S J_z,
\end{align}
and if $i$ is on an $\alpha$ chain:
\begin{align}
    B_i  = 2 S \left(\sin(\phi)^2\,J_x + \cos(\phi)^2\,J_z\right) + \cos(\phi-\theta)\, h.
\end{align}
\label{eq:app-rspt-B}
\end{subequations}

\subsection{Degeneracy lifting}

We are interested in the lowest order quantum correction that lifts the classical degeneracy. For that, the energy correction must be proportional to the product of as least two chains $\eta_i \eta_j$, which is only possible if the corresponding cluster connects two chains. 

\begin{figure}
    \centering
    \includegraphics{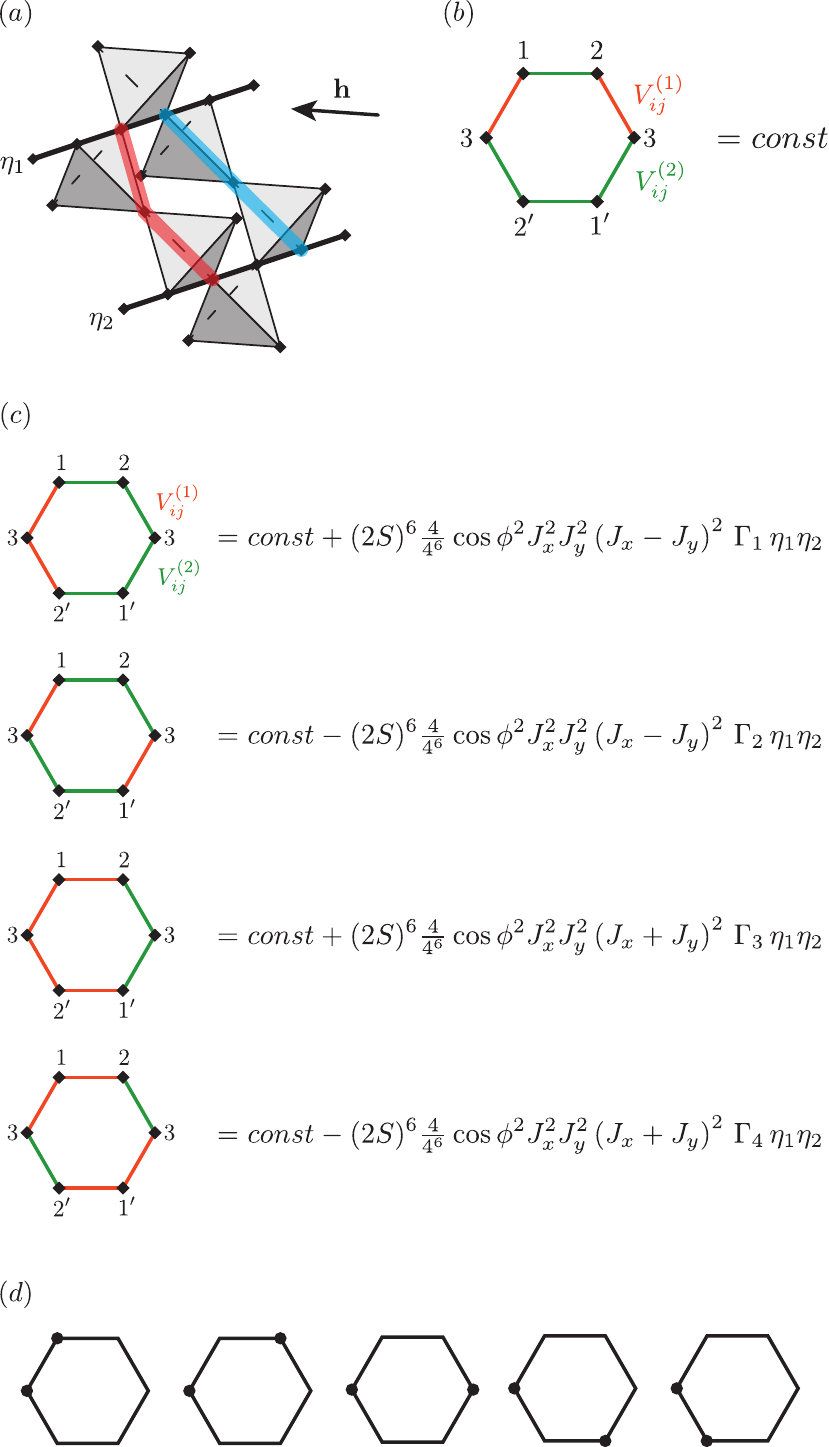}
    \caption{(a) The two distinct smallest clusters connecting two chains, marked in red and blue. Their lowest order contribution is of order 4 and they cancel out exactly. (b) A configuration of perturbation on a hexagon that does not contribute to the effective coupling. (c) The four inequivalent configurations that do contribute to $J_{\rm ch}$. (d) A possible sequence of intermediate states for the first configuration in (c). In (b) and (c), we label the sites with reference to the sublattice $\mathcal L_\nu$ they belong to, consistent with the hexagon indicated in green in \autoref{fig:ising} (a)}
    \label{fig:app-rspt}
\end{figure}

The two smallest clusters that connect two chains are shown in \autoref{fig:app-rspt} (a). They both have two links and their smallest contribution is of 4th order. Inspecting \autoref{eq:app1-matrix-elem-Vij}, we see that any term that is proportional to $\eta_c$ also has a factor $\sigma^{\beta}_i$. Hence, the two clusters marked in red and blue will cancel out exactly. This will be true for any cluster that has a `mirrored' cluster that connects to a site on the same chain but on a different sublattice.
Hence, only clusters of nontrivial topology, i.e. those that have noncontractible loops, will contribute to the effective chain couplings. The smallest such cluster on the pyrochlore lattice is a hexagon and there are two inequivalent ways a hexagon can connect two chains, shown in \autoref{fig:ising} (a). Note that for the same reasons, there will be no nonvanishing contribution from perturbation theory, that is proportional to only a single chain variable $\eta_i$. Any such term must be proportional to a respective $\sigma^{\beta}_i$ and has a mirrored term that gives the same contribution but with the opposite sign.

Evaluating the energy contributions of the two types of hexagons, we have to consider all possible distributions of $ V_{ij}^{(1)} $ and $V_{ij}^{(2)}$. However, not all of those contribute to the chain coupling. Consider for example the contribution from the cluster depicted in \autoref{fig:app-rspt} (b), which is [using \autoref{eq:app1-matrix-elem-Vij}]
\begin{align}
    \delta E^{(6)} \propto& \left(\cos\phi J_x + \eta_1 J_y \right) \left(\cos\phi J_x - \eta_1 J_y \right) \nonumber\\ 
    &\left(\cos\phi J_x + \eta_2 J_y \right) \left(\cos\phi J_x - \eta_2 J_y \right)(J_x - J_y)^2
\end{align}
and thus the dependence on $\eta_i$ drops out completely.

Keeping only those configurations which do depend on the chain configuration $\{\eta_i\}$, there are four inequivalent possibilities, shown in \autoref{fig:app-rspt} (c). Each of them have different possible sequences of intermediate states $\{\psi_i\}$, which must be enumerated to evaluate the denominator of \autoref{eq:app1-eval}. We show one possible such sequence, for the first configuration in \autoref{fig:app-rspt} (c), in \autoref{fig:app-rspt} (d), where the excited states are shown as dots on the sites.

Finally, the effective coupling $J_{\rm ch}$ is given as the sum of the four inequivalent contributions depicted in \autoref{fig:app-rspt} (c)
\begin{align}
    J_{\mathrm{ch}} =& \frac{4(2S)^6}{4^6} \cos(\phi)^2 J_x^2 J_y^2 \bigl( (J_x - J_y)^2 (\Gamma_1 - \Gamma_2)  \nonumber\\
    &+ (J_x + J_y)^2 (\Gamma_3 - \Gamma_4) \bigr),
\end{align}
where we have defined factors $\Gamma_i$, stemming from the denominator of \autoref{eq:app1-eval}
\begin{align}
    \Gamma_{\mathrm{config}} := -\sum_{\{\psi_i\}} \frac{1}{(E_0 - E_{\psi_1}) \dots (E_0 - E_{\psi_{p-1}})}
\end{align}
Note that since all factors in the denominator are negative and there are an odd number of factors for each closed loop on the pyrochlore lattice, $\Gamma_{\rm config}>0$.

Evaluating the $\Gamma$ factors, one finds that $\Gamma_3 = \Gamma_4$ and hence 
\begin{subequations}
\begin{align}
    J_{\mathrm{ch}}^z =& - S J_z \cos(\phi)^2 \frac{J_x^2 J_y^2}{J_z^4} \frac{\left( J_x - J_y\right)^2}{J_z^2} \frac{1}{16}  \Gamma_z(h),
\end{align}
where $\Gamma_z(h) := S^5 J_z^5 (\Gamma_2 - \Gamma_1)$ is defined as a dimensionless factor controlling the asymptotic field dependence of the couplings and we have introduced a superscript `$z$' to denote that the above result is true in the ${\rm Chain}_Z$ phase. 

\begin{figure}
    \centering
    \includegraphics{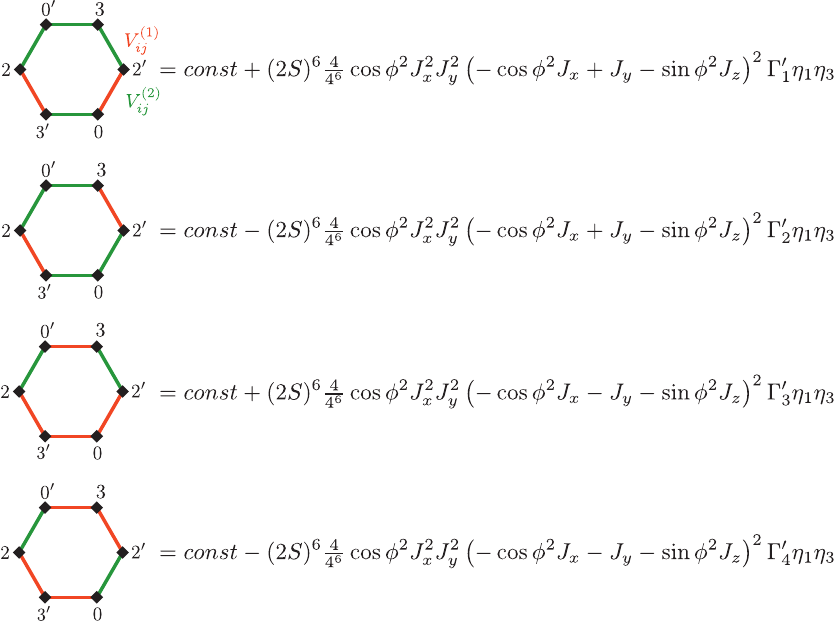}
    \caption{The four inequivalent configurations of perturbations on a hexagon contributing to $J_{\rm ch}'$. We label the sites with reference to the sublattice $\mathcal L_\nu$ they belong to, consistent with the hexagon indicated in pink in \autoref{fig:ising} (a)}
    \label{fig:app-rspt-2}
\end{figure}

Computing $J_{\mathrm{ch}}'$ follows exactly the same procedure. Since the sites on the hexagon, that is the leading-order contribution to $J_{\mathrm{ch}}'$ are on different sublattices, also configurations of perturbation on the hexagon that contribute will be different. The four inequivalent configurations that contribute to $J_{\mathrm{ch}}^{z\prime}$, together with their contributions are shown in \autoref{fig:app-rspt-2}. Summing those up yields 
\begin{align}
    J_{\mathrm{ch}}^{z\prime} =& S J_z \cos\phi^4\frac{J_x^2 J_y^3}{J_z^5} \left( J_x/J_z + \tan\phi^2 \right) \frac{2}{16}  \Gamma_z'(h).
\end{align}
\label{eq:effJ-rspt}
\end{subequations}
This completes the derivation of the results stated in the main text. For completeness, we give the $\Gamma$-factors explicitly:
\begin{subequations}
\begin{align}
    \frac{\Gamma_z(h)}{(SJ_z)^5} =& \frac{2}{B^\beta_z (B^\alpha_z + B^\beta_z)^4} 
        + \frac{2}{(B^\beta_z)^2 (B^\alpha_z + B^\beta_z)^3} \nonumber\\
        &+ \frac{1}{(B^\beta_z)^3 (B^\alpha_z + B^\beta_z)^2}, \\
    \frac{\Gamma'_z(h)}{(SJ_z)^5} =& \frac{4}{B^\beta_z (B^\alpha_z + B^\beta_z)^4},
\end{align}
\label{eq:app1-gamma}
\end{subequations}
where $B_z^\alpha$ and $B_z^\beta$ are the local fields on the $\alpha$ and $\beta$ chains respectively, as defined in \autoref{eq:app-rspt-B}.
Note that since the local fields $B^\alpha_\lambda$ and $B^\beta_\lambda$ are always positive, so are $\Gamma_z$ and $\Gamma_z'$

In the ${\rm Chain}_X$ phase, the ground state as defined in \autoref{eq:app1-basis} will be slightly different 
\begin{subequations}
\begin{align}
    \unitvec w_i =
    \begin{cases}
        \left(\sigma_i^{\alpha}\sin\phi, 0, \sigma_i^{\alpha}\cos\phi\right) & \text{$i\in\alpha$ chain}\\
        \left(\sigma_i^{\beta}\eta_c, 0, 0 \right) & \text{$i\in\beta$ chain}
    \end{cases},
\end{align}
and
\begin{align}
    \vec c_i = \frac{1}{2}
    \begin{cases}
        \left(\sigma_i^{\alpha}\cos\phi, i, -\sigma_i^{\alpha}\sin\phi\right) & \text{$i\in\alpha$ chain}\\
        \left(0, i\sigma_i^{\beta}\eta_c, -1 \right) & \text{$i\in\beta$ chain}
    \end{cases}.
\end{align}
\label{eq:app1-basis-x}
\end{subequations}
Note the sign change in the definition of $\vec c_i$ on $\beta$ chains that is necessary so the basis stays right-handed. Since $\vec J$ is diagonal, the above change can be accounted for by simply changing $J_x \leftrightarrow J_z$ and $\cos\phi \leftrightarrow \sin\phi$ in \autoref{eq:effJ-rspt} and hence
\begin{subequations}
\begin{align}
    J_{\mathrm{ch}}^x =& - S J_x \sin(\phi)^2 \frac{J_z^2 J_y^2}{J_x^4} \frac{\left( J_z - J_y\right)^2}{J_x^2} \frac{1}{16} \Gamma_x(h), \\
    J_{\mathrm{ch}}^{x\prime} =& S J_x \sin(\phi)^4\frac{J_x^2 J_y^3}{J_x^5} \left( J_z/J_x + \mathrm{cotan}\,\phi^2 \right) \frac{2}{16} \Gamma_x'(h).
\end{align}
\label{eq:effJ-x-rspt}
\end{subequations}
where $\Gamma_x^{(\prime)}$ is obtained from $\Gamma_z^{(\prime)}$ [\autoref{eq:app1-gamma}] by substituting $B_z^\alpha$ and $B_z^\beta$ by $B_x^\alpha$ and $B_x^\beta$ respectively.

\subsection{\texorpdfstring{Chain$_Y$}{Chain-Y} phase}

We now turn to calculate the effective couplings in the ${\rm Chain}_Y$ phase, which is done mostly by the same procedure as used for the ${\rm Chain}_{X/Z}$ phases. 
As before, we start by defining a basis where the $\unitvec w_i$ are given by the classical ground state configuration
\begin{subequations}
\begin{align}
    \unitvec w_i =
    \begin{cases}
        \left(\sigma_i^{\alpha}\sin\phi, 0, \sigma_i^{\alpha}\cos\phi\right) & \text{$i\in\alpha$ chain}\\
        \left(0, \sigma_i^{\beta}\eta_c, 0 \right) & \text{$i\in\beta$ chain}
    \end{cases},
\end{align}
and the $\vec c_i$ complete the basis
\begin{align}
    \vec c_i = \frac{1}{2}
    \begin{cases}
        \left(\sigma_i^{\alpha}\cos\phi, i, -\sigma_i^{\alpha}\sin\phi\right) & \text{$i\in\alpha$ chain}\\
        \left(i, 0, \sigma_i^{\beta}\eta_c \right) & \text{$i\in\beta$ chain}
    \end{cases}.
\end{align}
\label{eq:app1-basis-y}
\end{subequations}
Substituting the above into \autoref{eq:app1-V} yields if $i,j$ are both on a $\beta$ chain
\begin{subequations}
\begin{align}
    V_{ij}^{(1)} &= -\frac{1}{4} \left( J_x + J_z\right) S_i^+ S_j^+ + h.c.,\\
    V_{ij}^{(2)} &= \frac{1}{4} \left( J_x - J_z\right) S_i^+ S_j^- + h.c.,
\end{align}
in contrast, if $i,j$ are both on an $\alpha-$chain
\begin{align}
    V_{ij}^{(1)} &= -\frac{1}{4} \left(\cos\phi^2\,J_x + J_y + \sin\phi^2\,J_z\right) S_i^+ S_j^+ + h.c., \\
    V_{ij}^{(2)} &= -\frac{1}{4} \left(\cos\phi^2\,J_x - J_y + \sin\phi^2\,J_z\right) S_i^+ S_j^- + h.c., 
\end{align}
and finally if $i$ is on a $\beta-$chain and $j$ is on an $\alpha-$chain
\begin{align}
    V_{ij}^{(1)} &= -\frac{1}{4} \sigma^{\alpha}_j \left( i\cos\phi\, J_x + \sigma^{\beta}_i\eta_c \sin\phi\, J_z\right) S_i^+ S_j^+ + h.c., \\
    V_{ij}^{(2)} &= -\frac{1}{4} \sigma^{\alpha}_j \left( i\cos\phi\, J_x + \sigma^{\beta}_i\eta_c \sin\phi\, J_z\right) S_i^+ S_j^- + h.c..
\end{align}
\label{eq:app1-matrix-elem-Vij-y}
\end{subequations}
Note that in the last case, the matrix elements are now complex.

\subsubsection{Effective coupling \texorpdfstring{$J_{\rm ch}^{y}$}{Jch-y}}

We now turn to identify the degeneracy-breaking perturbations. In the case of the ${\rm Chain}_{Z/X}$ phases those are given by linked clusters of nontrivial topology. Note that the argument for this relied solely on the fact that any Ising chain variable $\eta_c$ comes also with a factor $\sigma^\beta_i$, which is clearly still the case. Hence we can again restrict the discussion to the two kinds of hexagons as long as we are interested only in leading-order corrections. 

\begin{figure}
    \centering
    \includegraphics{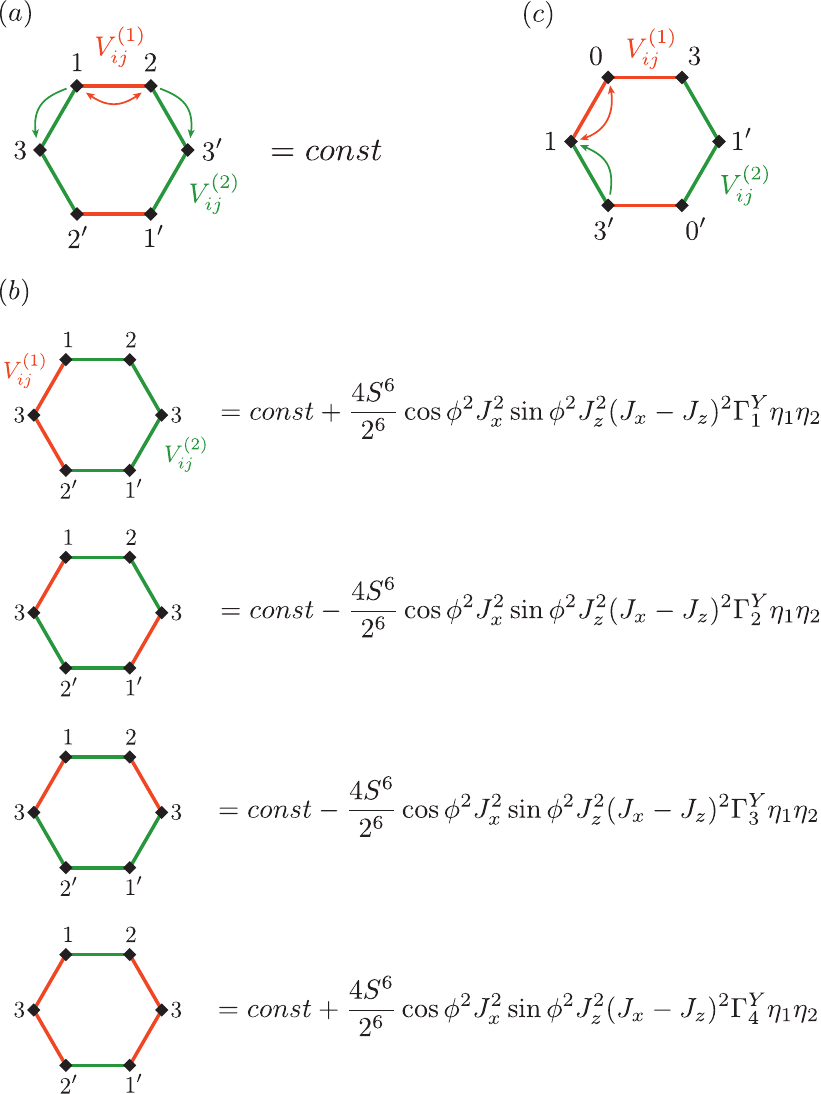}
    \caption{Diagrams arising in RSPT in the ${\rm Chain}_Y$ phase. (a) Example of a configuration that does not contribute, with one possible direction of spin flip hopping indicated. Whenever there is a double spin flip acting on the $\beta$ chain (site 1, 2), the two perturbations on edge $(1,2)$ and $(2,3')$ will have to act in the same way (i.e. both with $S^+$ or both with $S^-$) on site $1$ and $2$ respectively, leading to a constant contribution.
    (b) The four perturbations contributing to $J_{\rm ch}^{y}$. Note that the product of matrix elements does not depend on the order of application of the perturbations, so that the $\Gamma$ factors can be computed as before. (c) A distribution of perturbations on the hexagon whose product of matrix elements yields a contribution to $J_{\rm ch}^{y\prime}$. However, the diagram is not valid since there is no possible order in which, starting from the ground state, it yields the ground state also as a final state. For spin-$1/2$, there are actually no allowed diagrams using only $V_{ij}^{(1)}$, $V_{ij}^{(2)}$ and contributing to $J_{\rm ch}'$, even at higher orders in RSPT.}
    \label{fig:app-rspt-y}
\end{figure}

As before, not all distributions of the perturbations $V_{ij}^{(1)}$ and $V_{ij}^{(2)}$ yield an energy correction that is proportional to the two Ising variables $\eta_1$, $\eta_2$ on the hexagon. For that, first consider the case of $J_{\rm ch}^y$, which is generated by the hexagon marked in green in \autoref{fig:ising} (a). In this case, proportionality to a chain variable $\eta_1$ comes from a product of two perturbations connecting that chain to a polarized site, where the factor $\sigma^\beta_i$ is different for both. Inspecting \autoref{eq:app1-matrix-elem-Vij-y}, we see that the two perturbations have to act differently on the two $\beta$ chain sites in order for the resulting energy correction to be proportional to $\eta_1$. Note that this rule already excludes the possibility of having $V^{(1)}_{ij}$ act on the $\beta$ chain bonds [see \autoref{fig:app-rspt-y} (a)] and hence $J_{\rm ch}\propto (J_x-J_z)^2$ in accordance with the LSWT result (see \autoref{fig:obd-phases-1} and \ref{fig:obd-phases-2}).
Fixing the perturbations acting on $\beta$ chains to spin-flip hopping, there are four diagrams contributing to $J_{\rm ch}^y$ shown in \autoref{fig:app-rspt-y} (b). Summing up their contributions yields
\begin{align}
    J_{\rm ch}^y &= -\frac{1}{64} S J_y \sin(2\phi)^2 \frac{J_x^2 J_z^2 (J_x - J_z)^2}{J_y^6} \Gamma_y(h), \\
    \Gamma_y &= S^5 J_y^5 \left( 2\Gamma_2^Y + 2\Gamma_3^Y - 2\Gamma_1^Y - \Gamma_4^Y\right),
\end{align}
where the $\Gamma^Y_i$ are computed as before by enumerating all possible intermediate states and $\Gamma_y$ is given by
\begin{align}
    \frac{\Gamma_y}{(SJ_y)^5} =& \frac{2}{(B_y^\beta)^3 (B_y^\alpha + B_y^\beta)^2} - \frac{2}{(B_y^\alpha + 2B_y^\beta)^3(B_y^\alpha + B_y^\beta)^2} \nonumber\\
        &+ \frac{4}{(B_y^\beta)^2 (B_y^\alpha + B_y^\beta)^3} - \frac{4}{(B_y^\alpha + 2B_y^\beta)^2 (B_y^\alpha + B_y^\beta)^3} \nonumber\\
        &+ \frac{2}{B_y^\beta (B_y^\alpha + B_y^\beta)^4} - \frac{2}{(B_y^\alpha + 2B_y^\beta)(B_y^\alpha + B_y^\beta)^4},
\end{align}
where $B_y^\alpha$ and $B_y^\beta$ are the local fields on the $\alpha$  and $\beta$ chains respectively in the ${\rm Chain}_Y$ phase, obtained from \autoref{eq:app-rspt-B} by replacing $J_z$ by $J_y$. Note that since the local fields are always positive, it is clear that $\Gamma_y>0$ and hence the effective chain coupling $J_{\rm ch}^y<0$ is ferromagnetic for all fields and exchange parameters.

\subsubsection{Effective coupling \texorpdfstring{$J_{\rm ch}^{y\prime}$}{Jch-y'}}

We now turn to the second effective exchange coupling, $J_{\rm ch}^{y\prime}$, which is generated by the hexagon marked in pink in \autoref{fig:ising} (a). On the level of RSPT diagrams, this amounts to a different distribution of polarized and unpolarized sites on the hexagon as indicated in \autoref{fig:app-rspt-y} (c). Turning to identify the degeneracy breaking perturbations, we straightforwardly arrive at a similar rule as in the case of $J_{\rm ch}^y$. However, since the two perturbations connecting the site on the $\beta$ chain now connect to the same site, and because $\sigma_j^\alpha$ appears just as an overall sign in the matrix elements [\autoref{eq:app1-matrix-elem-Vij-y}], the two perturbations now have to act in the same way (i.e. either both with $S^+$ of both with $S^-$) on the site on the $\beta$ chains as indicated in \autoref{fig:app-rspt-y} (c). 
However, this is not possible at sixth order since there can be no double occupation of excitations on any site. 
For the same reason, for $S = 1/2$ any diagram using only $V_{ij}^{(1)}$, $V_{ij}^{(2)}$ cannot yield a finite contribution to $J_{\rm ch}^{y\prime}$, even at higher orders.
Since in the ${\rm Chain}_Y$ phase $V_{ij}^{(3)}=0$ if $i$ on a $\beta$ chain and $j$ on an $\alpha$ chain (or the other way around), a finite contribution to $J_{\rm ch}'$ can only arise from a perturbation including $V_{ij}^{(4)}$. Such a term occurs at seventh order in RSPT and competes with contributions from higher orders to $J_{\rm ch}$, even when the sixth order contribution to $J_{\rm ch}$ vanishes.

Since in dipolar-octupolar pyrochlores the pseudospin always forms a doublet, we conclude that $J_{\rm ch}^{y\prime}$ is strongly suppressed for these compounds. Taken together with the fact that $J_{\rm ch}^y < 0$, we conclude that quantum fluctuations should mostly or even always select a ferromagnetic chain configuration in the ${\rm Chain}_Y$ phase. 
LSWT (see \autoref{fig:obd-phases-1} and \ref{fig:obd-phases-2}) reproduces the overall dominance of ferromagnetic ordering at high fields, but predicts a zigzag ground state for some exchange parameters at low fields. We ascribe this to the fact that LSWT does not capture the hardcore nature of magnons at $S=1/2$ since $V_{ij}^{(4)}$ corresponds to cubic magnon terms in spin wave theory.

Notwithstanding the ambiguity of the two methods at low fields, at higher fields both LSWT and RSPT calculations agree that ferromagnetic chain configurations will dominate the ${\rm Chain}_Y$ phase [see \autoref{fig:obd-phases-1} (b) and \ref{fig:obd-phases-2} (b)].

\section{Linear spin wave theory\label{app:lswt}}

In this appendix, we provide details about the linear spin wave theory calculations presented in the main text.
We start from a system consisting of $L$ unit cells $U_l$ ($l=0\dots L-1$), repeated along one dimension. Each unit cell has $M$ sites at positions $l \vec a + \vec r_i$ ($i=0\dots M-1$). 
The general nearest-neighbor exchange Hamiltonian of such a system reads
\begin{align}
    \mathcal H =& \frac{1}{2} \sum_l \sum_{i,j\in U_l} \vec S_i \mathcal J_{ij} \vec S_j
        + \sum_l \sum_{i\in U_l} \sum_{j\in U_{l+1}} \vec S_i \mathcal K_{ij} \vec S_j \nonumber\\
        &- \sum_l \sum_{i \in U_l} \vec h_i \cdot \vec S_i,
\end{align}
where $\mathcal J_{ij}$ and  $\mathcal K_{ij}$ parametrize the exchange within and between unit cells respectively.

To derive the linear spin wave Hamiltonian, we first write the spins in a local basis ${\bf u}_i,{\bf v}_i,{\bf w}_i$, chosen such that in the classical ground state all spins align with ${\bf w}_i$
\begin{align}
    \vec S_i = S_i^{u} \vec u_i + S_i^{v} \vec v_i + S_i^{w} \vec w_i,
\end{align}
and define raising and lowering operators via the (linear) Holstein-Primakoff transformation
\begin{subequations}
\begin{align}
    S_i^{u} &\approx \sqrt{S/2} \left(a_i + a_i^\dagger \right), \\
    S_i^{v} &\approx \sqrt{S/2} \left(a_i - a_i^\dagger \right), \\
    S_i^{w} &= S - a_i^\dagger a_i.
    \label{eq:lswt-S}
\end{align}
\end{subequations}
Then, we perform a one-dimensional Fourier transform on sublattice $i$
\begin{align}
    a^\dagger_i(\vec q) = \frac{1}{\sqrt L} \sum_l \exp(-i \vec q \cdot (l \vec a  + \vec r_i)),
\end{align}
where the momentum takes values $\vec q = 2\pi \vec a k/ \abs{\vec a}^2 L$ ($k=0\dots L-1$). Using the above, we arrive at
\begin{align}
    \mathcal H =& E_0 + \mathcal H_1 + \mathcal H_2,
\end{align}
where $E_0$ is the classical ground state energy, $\mathcal H_1$ contains terms linear in the operators $a_i$, $a_i^\dagger$ and vanishes if $\vec S_i^{(0)} = S\unitvec w_i$ is a classical ground state. Finally, $\mathcal H_2$ is given by
\begin{subequations}
\begin{align}
    \mathcal H_2 &= \frac{1}{2} \sum_{\vec q} \vec A(\vec q)^\dagger \mathcal X(\vec q) \vec A(\vec q), \\
    \vec A^\dagger(\vec q)  &= (a_0^\dagger(\vec q), \dots, a_{M-1}^\dagger(\vec q), a_0(-\vec q), \dots), \\
    \mathcal X(\vec q) &= S \mqty(X^{(11)}(\vec q) & X^{(12)}(\vec q) \\ X^{(21)}(\vec q) & X^{(22)}(\vec q)), 
\end{align}

\begin{widetext}
\begin{align}
    X_{ij}^{(11)}(\vec q) &= e^{-i \vec q \cdot \vec r_{ij}} \Biggl(
            \vec c_i \mathcal J_{ij} \vec c_j^*
            + e^{-i \vec q \cdot \vec a} \vec c_i \mathcal K_{ij} \vec c_j^*
            + e^{i \vec q \cdot \vec a} \vec c_j^* \mathcal K_{ji} \vec c_i \nonumber\\
            &\phantom{=e^{-i \vec q \cdot \vec r_{ij}} \Biggl(}
                - \delta_{ij} \left(
                    \sum_{l=0\dots M-1} \left(
                        \vec w_i \mathcal J_{il} \vec w_l
                        + \vec w_i \mathcal K_{il} \vec w_l
                        + \vec w_l \mathcal K_{li} \vec w_i
                    \right)
                    - \frac{1}{S} \vec h_i \cdot \vec w_i
                \right)
         \Biggr), \\
    X_{ij}^{(12)}(\vec q) &= e^{-i \vec q \cdot \vec r_{ij}} \left(
        \vec c_i \mathcal J_{ij} \vec c_j
        + e^{-i \vec q \cdot \vec a} \vec c_i \mathcal K_{ij} \vec c_j
        + e^{i \vec q \cdot \vec a} \vec c_j \mathcal K_{ji} \vec c_i
        \right), \\
    X_{ij}^{(21)}(\vec q) &= e^{-i \vec q \cdot \vec r_{ij}} \left(
        \vec c_i^* \mathcal J_{ij} \vec c_j^*
        + e^{-i \vec q \cdot \vec a} \vec c_i^* \mathcal K_{ij} \vec c_j^*
        + e^{i \vec q \cdot \vec a} \vec c_j^* \mathcal K_{ji} \vec c_i^*  \right), \\
    X_{ij}^{(22)}(\vec q) &= e^{-i \vec q \cdot \vec r_{ij}} \Biggl(
            \vec c_i^* \mathcal J_{ij} \vec c_j
            + e^{-i \vec q \cdot \vec a} \vec c_i^* \mathcal K_{ij} \vec c_j
            + e^{i \vec q \cdot \vec a} \vec c_j \mathcal K_{ji} \vec c_i^* \nonumber\\
            &\phantom{=e^{-i \vec q \cdot \vec r_{ij}} \Biggl(}
                - \delta_{ij} \left(
                    \sum_{l=0\dots M-1} \left(
                        \vec w_i \mathcal J_{il} \vec w_l
                        + \vec w_i \mathcal K_{il} \vec w_l
                        + \vec w_l \mathcal K_{li} \vec w_i
                    \right)
                    - \frac{1}{S} \vec h_i \cdot \vec w_i
                \right)
         \Biggr),
\end{align}
\end{widetext}
\label{app2-lswt-hamiltonian}
\end{subequations}
where the $X^{(nm)}(\vec q)$ are $M\times M$ matrices. The sum goes over $\vec q = 2\pi \vec a k/ \abs{\vec a}^2 L$ ($k=0\dots L-1$) and $\vec a$ is the translation vector between neighboring unit cells and $\vec r_{ij} = \vec r_j - \vec r_i$ is the difference vector between two sites within the unit same cell.

\autoref{app2-lswt-hamiltonian} is quadratic and thus can be diagonalized using a Bogoliubov transformation. Explicitly, the magnon dispersion is obtained by diagonalizing $\sigma \mathcal X(\vec q)$ where
\begin{equation}
    \sigma = \mqty(\id_{M\times M} & 0 \\ 0 & -\id_{M\times M}).
    \label{eq:app2-sigma}
\end{equation}
The matrix $\sigma \mathcal X(\vec q)$ has eigenvalues $\omega_\nu(\vec q)$, $-\omega_\nu^*(\vec q)$, $\nu=0\dots M-1$.

The magnon spectrum $\omega_\nu(\vec q)$ depends on the classical ground state $\{\vec S_i^{(0}\}$ used in the expansion above. The ground state selection by quantum fluctuations can then be quantified using the zero point energy
\begin{align}
    \mathcal E_0 \left( \{\vec S_i^{(0}\} \right) &= \lim_{L\to\infty} \frac{1}{L}\bra{0} H_2 \ket{0} \\
        &= \frac{\abs{\vec a}}{4\pi} \sum_\nu \int_{0}^{2\pi/\abs{\vec a}} \dd k\, \omega_\nu \left(k \vec a / \abs{\vec a}\right).
    \label{eq:app1-zpe-general}
\end{align}

\subsection{Details of the numerical calculations}

Here, we provide some details of the computation of the numerical phase diagram in \autoref{fig:obd-phases-1} and \ref{fig:obd-phases-2}. For a discussion of the different phases, we refer to the main text [\autoref{sec:obd-phases}].

As mentioned already in the main text, the LSWT theory calculations are performed on a semi-infinite ``tube'' of the pyrochlore lattice, that is an effective $W\times W$ triangular lattice of $\beta$ chains. The infinite direction is thus $[1\bar10]$ [$\vec a = (2, -2, 0)$] and the cluster has $M = 4 W^2$ sites per unit cell. 
The zero point energy is then given by
\begin{align}
    \mathcal E_0 \left( \{\vec S_i^{(0}\} \right) &= \frac{1}{\pi} \sum_\nu \int_0^{\pi/2} \dd k\, \omega_\nu \left((k, -k, 0)\right),
    \label{eq:app1-zpe}
\end{align}
where we used that the dispersion is symmetric around $k=0$.

For a fixed set of exchange couplings $\vec J$ and field $\vec h$, we diagonalize the resulting $2M\times 2M$ matrix numerically for 50 values of $\vec q$ and compute the integral of the dispersion numerically using the standard Simpson's quadrature rule. Since relative differences between different zero point energies are very small, we verified that using 50 points, the integral is converged up to double precision. For $W=8$, computing the zero point energy of a single classical ground state for fixed parameters takes about $10$ seconds.

The number of distinct configurations of the Ising variables $\{\eta_i\}$ grows exponentially with the system size $W$. The last system size for which it is feasible to enumerate all possible configurations is $W=4$, for which there are $2^{W^2}=65 536$. This can be reduced to $674$ inequivalent chain configurations which are not related by symmetries. Computing the zero point energies [\autoref{eq:app1-zpe}] for all these states in linear spin wave theory yields the phase boundary between the ferromagnetic and the zigzag state, as well as the gap $\Delta$ between the two. 
However, at this system size, LSWT is not able to correctly capture the splitting between zigzag states. This is because the unit cell of the two chosen ground states, the kinked and the striped states is of size $W=4$. If we would simulate the effective triangular lattice Ising model directly, a system of the size of that unit cell would be sufficient to capture the selection process.

However, since we actually consider the whole pyrochlore lattice and wish to capture the
fluctuations which generate the interactions between the chains, we need a cluster which is at least twice as large to suppress the influence of virtual processes which cross the entire cluster via the periodic boundary conditions.

In order to obtain the fully resolved phase diagram shown in \autoref{fig:obd-phases-1} and \ref{fig:obd-phases-2}, we use a system with size $W=8$. For each set of parameters, we compute the zero point energy of all possible ground states of the effective model, that are the ferromagnet, the staggered state, and all inequivalent zigzag states. The latter is possible because the total number of zigzag states grows exponentially only in the linear system size, hence there are $2^W=256$ zigzag states in total. The number can be further reduced to 12 zigzag states not related by symmetry. To summarize, there are 14 possible inequivalent ground states of the effective model for $W=8$.

\begin{figure}
    \centering
    \includegraphics{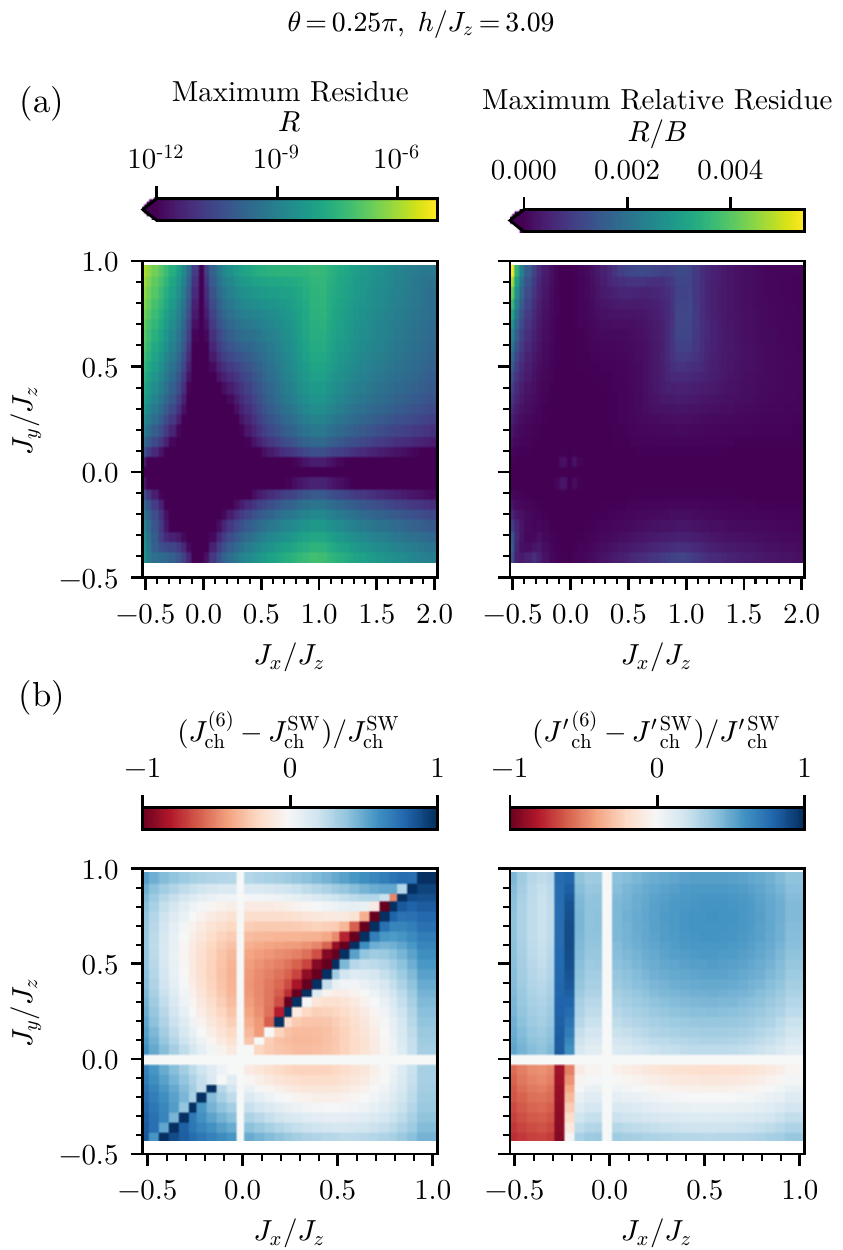}
    \caption{Benchmarking the effective model using linear spin wave theory (LSWT) using the $14$ ground states of the effective model and $1000$ random chain configurations. (a) Maximum residue $R$ between a zero point energy (per site) predicted by the effective model and computed by LSWT. We give its absolute value as well as divided by the total bandwidth of zero point energies $B$, which is a scale given by the effective nearest-neighbor chain couplings. (b) Comparison between the nearest-neighbor effective chain couplings as calculated from 6th real-space perturbation theory ($J_{\rm ch}^{(\prime)(6)}$) and as obtained from LSWT using \autoref{eq:eff-j-lswt} ($J_{\rm ch}^{(\prime)\rm SW}$). }
    \label{fig:eff-model-res}
\end{figure}

To benchmark the validity of the effective triangular lattice Ising model we calculate the zero-point energy of 1000 random states. We show the results of this benchmarking in  \autoref{fig:eff-model-res}. In (a) we compare the zero point energies obtained by LSWT to those computed from an effective model with effective couplings fitted using three states [\autoref{eq:eff-j-lswt}]. In (b), we compare the effective chain couplings obtained from 6th order RSPT ($J_{\rm ch}^{(\prime)(6)}$) to those computed from LSWT ($J_{\rm ch}^{(\prime)\rm SW}$).

Finally, to assess finite size effects we also simulate a system with $W=16$ and verify that the numerical values of $\Delta$ and $B$ are independent of the size.

\subsection{Survival of full classical degeneracy when \texorpdfstring{$J_x=0$}{Jx=0} or \texorpdfstring{$J_y=0$}{Jy=0} in the dipolar chain phases}

In this appendix, we prove the survival of the full classical degeneracy, within LSWT, in the singular cases of $J_x=0$ or $J_y=0$. To this end, we relate the flipping of single $\beta$ chain in the system to a unitary transformation on the matrix $\sigma \mathcal X(\vec q)$. This already implies that the spectrum and hence the zero point energy of the system does not depend on the chain configuration $\{\eta_i\}$ and hence the full classical degeneracy is preserved even in the presence of quantum zero-point fluctuations.

In the dipolar chain phase i.e. in either ${\rm Chain}_X$ or ${\rm Chain}_Z$, all classical ground states are periodic in the $[1\bar10]$ direction with a unit cell of four sites per $\beta$ chain in the system. Two sites of those are on the $\beta$ chain and the other two are on $\alpha$ chains.
We now consider the case of a general (nonzero) matrix element of $\mathcal X(\vec q)$ as given in \autoref{app2-lswt-hamiltonian}. Flipping a single chain variable $\eta_c \to -\eta_c$ amounts to a change in a maximum of $18$ matrix elements in each of the four submatrices $X^{(mn)}$ of $\mathcal X(\vec q)$. These matrix elements can take two forms.

In the following, we assume for simplicity to be in the ${\rm Chain}_Z$ phase, but the same argument applies in the ${\rm Chain}_X$ phase. First, if $i,j$ are on the same $\beta$ chain, the matrix elements are given by
\begin{subequations}
\begin{align}
    X^{(11)}_{ij}(\vec q) &= X^{(22)}_{ij}(\vec q) = \cos(\pi k /L) \left(J_x - J_y\right), \\
    X^{(12)}_{ij}(\vec q) &= X^{(21)}_{ij}(\vec q) = \cos(\pi k /L) \left(J_x + J_y\right),
\end{align}
\end{subequations}
where we used the definition of $\vec c_i$ from \autoref{eq:app1-basis}. The above is independent of the Ising variable $\eta_c$ of the chain and hence there are only $16$ matrix elements left that might change as the chain is flipped. 
The other sixteen matrix elements are all of the form 
\begin{subequations}
\begin{align}
    X^{(11)}_{ij}(\vec q) &= X^{(22)}_{ij}(\vec q) \nonumber\\
        &= \frac{1}{2} \exp(\pm i\vec q \cdot \vec a/4)
        \left( \sigma_j^\alpha J_x \cos\phi + \sigma_i^\beta\eta_c J_y \right), \\
    X^{(12)}_{ij}(\vec q) &= X^{(21)}_{ij}(\vec q) \nonumber\\
        &= -\frac{1}{2} \exp(\pm i\vec q \cdot \vec a/4)
        \left( \sigma_j^\alpha J_x \cos\phi - \sigma_i^\beta\eta_c J_y \right),
\end{align}
\label{eq:X-matrix-elem}
\end{subequations}
where $i$ is a site on a $\beta$ chain, $j$ is a site on an $\alpha$ chain and $\sigma_i^\alpha$, $\sigma_j^\beta$ are defined in \autoref{eq:app1-sigma}. The sign of the site-dependent phase $\pm i \vec q \cdot \vec a/4$ depends on the specific sites $i$ and $j$, but this explicit dependence does not matter for the argument presented here.
For $J_y = 0$, it is clear that the spin wave matrix $\mathcal X(\vec q)$ does not depend on the Ising variable $\eta_c$ and hence there is no ground state selection. For $J_x=0$ the argument is less obvious but the same is still true. In that case, all matrix elements that change when $\eta_c \to -\eta_c$ are proportional to $\eta_c$. Thus, flipping the chain is equivalent to a simple unitary transformation on $\sigma \mathcal X(\vec q)$ with the transformation matrix given by 
\begin{equation}
    U = \mathrm{diag}(1, \dots, -1, -1, \dots, -1,-1, \dots 1),
\end{equation}
where the $-1$ entries appear at the positions of the two sites on the flipped $\beta$ chain.

\section{Perturbative limit of the \texorpdfstring{${\rm Chain}_Y^*$}{Chain-Y*} phase \label{app:quantum-chain-y-star}}

In this appendix, we show that the ${\rm Chain}_Y^*$ phase can be derived also in degenerate perturbation theory, from the limit of the octupolar ice phase ${\rm SI}_Y$. 

For simplicity, we set the transverse coupling to zero and study the Hamiltonian
\begin{subequations}
\begin{align}
    \mathcal H &= \mathcal H_0 + V, \\
    \mathcal H_0 &= J_y \sum_{\expval{ij}} S_i^y S_j^y, \\ 
    V &= - h \sum_{i\in \mathcal L_0} \left(S_i^+ + S_i^- \right) + h \sum_{i\in \mathcal L_3} \left(S_i^+ + S_i^- \right).
\end{align}
\label{eq:appc-H}
\end{subequations}
The ground states of $\mathcal H_0$ are given by the octupolar spin-ice configurations, where $\sum_{i\in {\rm tet}}S_i^y = 0$ on every tetrahedron. 

The leading order correction appears at second order and consists of the perturbation $V_i$ acting on the same site twice
\begin{align}
    \Delta E_i^{(2)} &= \bra{\psi_0} V_i \left( E - H_0 \right)^{-1} V_i \ket{\psi_0} \\
        &= \frac{2Sh^2}{-2S J_y} = -\frac{h^2}{J_y}.
\end{align}
The above is just a constant correction to the unperturbed ground state energy $E_0$, that is it does not discriminate between ground states $\psi_0$, leaving the degeneracy unchanged. 

This changes at fourth order, where the only nonvanishing contribution consists of application of the perturbation $V_i$ to two neighboring sites. Note that the perturbation acts only on the $\alpha$ chains. In the following, we assume that $S=1/2$, however qualitatively the result is the same for other values of $S$. The energy correction is then given by
\begin{align}
    \Delta E_i^{(4)} &= 4 \times \frac{(2S)^2 h^4}{-(2SJ_y)^2 S J_y (3 + \sigma_i \sigma_j)} \\
        &= const + \frac{h^4}{2J_y^3} \sigma_i \sigma_j,
\end{align}
where $\sigma_i=\pm1$ denotes the configuration of the octupolar moment on site $i$ in the ground state $\psi_0$. The above expression now does discriminate between ground states $\psi_0$, favoring those states with a staggered configuration of octupolar moments on the $\alpha$ chains.

The ice rule $\sum_{i \in tet} S^y_i=0$ then forces all $\beta$ chains into a staggered configuration 
This leaves an Ising degree of freedom on each of the $\alpha$ chains and each of the $\beta$ chains, reproducing exactly the degeneracy structure of the classical ${\rm Chain}_Y^*$ phase.

\begin{figure}
    \centering
    \includegraphics{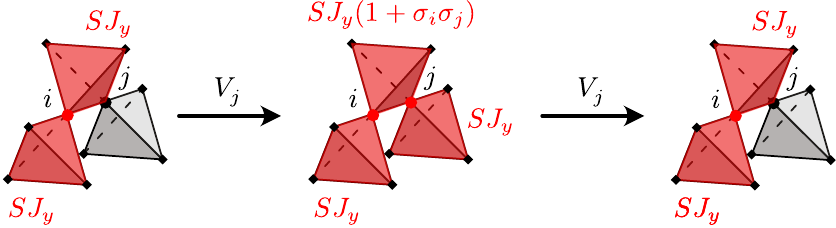}
    \caption{A possible sequence of intermediate states at fourth order degenerate perturbation theory. The full sequence applied here is $V_iV_jV_jV_i$. Tetrahedra on which the pseudospin two-in-two-out rule is violated are indicated in red, with energy penalties $\Delta E_{\rm tet}$ indicated next to them. Note that all valid orders of application of the $V_i$, $V_j$ will have the same set of intermediate energies since the ground state is not allowed as an intermediate state. }
    \label{fig:app-degen-perturb}
\end{figure}

The above result can be easily understood. First, since any valid sequence flips two spins twice, the numerator is given by $(2S)^2h^4$. 
A possible sequence of intermediate states is shown in \autoref{fig:app-degen-perturb} with contributions from each tetrahedron to the energy relative to the ground state indicated in red. Since the ground state is not allowed as an intermediate state, there are exactly three other possible sequences which all have the same sequence of intermediate energies. 
The first and last state have one of the two moments $S_i^y$ or $S_j^y$ flipped, and hence have an energy of $E-E_0=2SJ_y$.
The second intermediate state has both moments flipped and thus violates the spin-ice rule on three tetrahedra. Two of those include just one of the two sites and hence contribute also a term $2SJ_y$. The tetrahedron that shares the two sites however contributes an energy depending on the ground state configuration. If the two flipped sites where initially oriented antiparallel, then the spin ice rule is still fulfilled on this tetrahedron if both are flipped, while in the case of the two sites being parallel, the spin ice rule is now maximally violated. Altogether the energy of the second intermediate state can thus be written as $E-E_0=SJ_y(3+\sigma_i\sigma_j)$.

Adding transverse exchange ($J_x, J_z$) to \autoref{eq:appc-H} will allow the appearance of a ring exchange term in perturbation theory which will favour a U(1) QSL \cite{Shannon2012}.
The competition between this ring exchange and the diagonal term $\Delta E_{ij}^{(4)}$ will determine the low field phase boundary between the QSL and the ${\rm Chain}_Y^{\ast}$ phase.
Since the transition from QSL to ${\rm Chain}_Y^{\ast}$ does not require the condensation of spinons, but instead confines their dynamics it is a type of confinement transition.
This confinement transition should precede the Anderson-Higgs type transition from spinon condensation predicted in gauge Mean Field Theory calculations \cite{Li2017, Yao2020}, which generally requires $h\sim J_y$ in order to close the spinon gap.

\section{Effect of finite temperature\label{app:temperature}}

In this appendix, we study the effect of finite temperature of ground state selection. First we consider the low-temperature expansion of the classical free energy around different ground states and show that the first order correction is independent of the ground state. Second, we consider the free energy of the magnons at finite temperature.
We show that this leads to the same ground state selection
for all temperatures.
For temperatures below a crossover temperature $T_{\rm co}$, set by the spin wave gap, the strength of ground state selection is close to the zero temperature value. For  $T>T_{\rm co}$ the difference in free energy between different chain configurations is heavily suppressed. For parameters of interest (close to those estimated for Nd$_2$Zr$_2$O$_7$) this crossover temperature is given by $T_{\rm co} \approx 0.2 J_z$ (see \autoref{fig:obt-tempterature}).

\subsection{Low temperature expansion}

For the low temperature expansion, we choose as a starting point the same general one-dimensional Hamiltonian as for linear spin wave theory (\appref{app:lswt}).

That is a system consisting of $L$ unit cells $U_l$ ($l=0\dots L-1$), repeated along one dimension. Each unit cell has $M$ sites at positions $l \vec a + \vec r_i$ ($i=0\dots M-1$) and the Hamiltonian reads
\begin{align}
    \mathcal H =& \frac{1}{2} \sum_l \sum_{i,j\in U_l} \vec S_i \mathcal J_{ij} \vec S_j
        + \sum_l \sum_{i\in U_l} \sum_{j\in U_{l+1}} \vec S_i \mathcal K_{ij} \vec S_j \nonumber\\
        &- \sum_l \sum_{i \in U_l} \vec h_i \cdot \vec S_i,
        \label{eq:1d-H}
\end{align}
We again proceed by writing the spin variables ins a local basis $\vec u_i, \vec v_i, \vec w_i$ chosen such that in the (classical) ground state all spins align with $\vec w_i$. We can hence parametrize fluctuations around the ground state as
\begin{align}
    \vec S_i &= \sqrt S\, \delta u_i \vec u_i  +
        \sqrt S\, \delta v_i \vec v_i +
        \sqrt{S^2 - S (\delta u_i)^2 - S (\delta v_i)^2 } \vec w_i \nonumber\\
        &\approx \sqrt S\,\delta u_i \vec u_i +  
        \sqrt S\, \delta v_i \vec v_i +
        \left(S - \tfrac{1}{2}(\delta u_i)^2 - \tfrac{1}{2}(\delta v_i)^2\right) \vec w_i 
        \label{eq:low-T-S}
\end{align}
substituting the above into \autoref{eq:1d-H} up to quadratic order in the fluctuations $\delta u_i$, $\delta v_i$ yields
\begin{subequations}
\begin{align}
    \mathcal H = E_0 + \mathcal H_1 + \mathcal H_2
\end{align}
where $E_0$ is the ground state energy, $\mathcal H_1$ is linear in the fluctuations $\delta u_i$, $\delta v_i$ and vanishes if $\vec S^{(0)}_i = S \vec w_i$ is a classical ground state, and $\mathcal H_2$ is given by
\begin{align}
    \mathcal H_2 &= \frac{1}{2} \sum_{\vec q} \tilde{\vec u}(-\vec q)^T \mathfrak X(\vec q)  \tilde{\vec u}(\vec q), \\
    \tilde{\vec u}(\vec q)^T &= \left(\delta u_1(q), \delta u_2(q), \dots, \delta u_M(q), \delta v_1(q), \dots  \right).
\end{align}
\label{eq:app4-low-t-H}
\end{subequations}
Here, the matrix $\mathfrak X(\vec q)$ is related to the spin wave Hamiltonian $\mathcal X(\vec q)$ \autoref{app2-lswt-hamiltonian} derived in \appref{app:lswt} by a unitary transformation
\begin{subequations}
\begin{align}
    \mathfrak X(\vec q) &= F^\dagger \mathcal X(\vec q) F \\ 
    F &= \frac{1}{\sqrt{2}} \mqty(\id & \phantom{-}i\id \\ \id & -i\id)
\end{align}
\label{eq:x-cq-trafo}
\end{subequations}

The classical partition function can be calculated explicitly
\begin{align}
    Z &= \tr e^{-\beta \mathcal H} = (2\pi)^{-ML} \int \left(\prod_{\nu=1}^{2M} \dd \tilde u_\nu \right) e^{-\beta E_0} 
        e^{-\beta \mathcal H_2} \nonumber\\
    &= e^{-\beta E_0} \prod_{\vec q} \left(\det \beta \mathfrak X(\vec q) \right)^{\tfrac{1}{2}}
    \label{eq:app4-Z}
\end{align}
and the free energy up to linear order in temperature by
\begin{align}
    F &= -T \log Z \nonumber\\
    &= E_0 + \frac{T}{2} \sum_{\vec q} \log \det \mathfrak X(\vec q) - TML \log T.
\end{align}

Now, note that flipping a chain in all cases amounts to flipping $\vec w_i$ on two sites. To keep the basis right handed, one then also has to flip either $\vec u_i$ or $\vec v_i$ on those sites. 
In the expression for the partition function (\ref{eq:app4-Z}) however, assuming we flipped  $\vec u_i$, one can revert this change by the transformation of variables $\delta u_i \to -\delta u_i$. Since the sign of an even number of variables is flipped, this leaves the phase space volume element $\left(\prod_{\nu=1}^{2M} \dd \tilde u_\nu \right)$ invariant. Hence, the partition function $Z$ as well as the free energy $F$ are invariant under flipping any single chain variable $\eta_c$ up to second order in deviations from the ground state configuration. Generally the above argument implies that for a set of (locally) collinear ground states as considered here in all of the chain phases, there is no order-by-disorder when taking the purely classical limit.

\subsection{Free energy of magnons at finite temperature}

\begin{figure}
    \centering
    \includegraphics{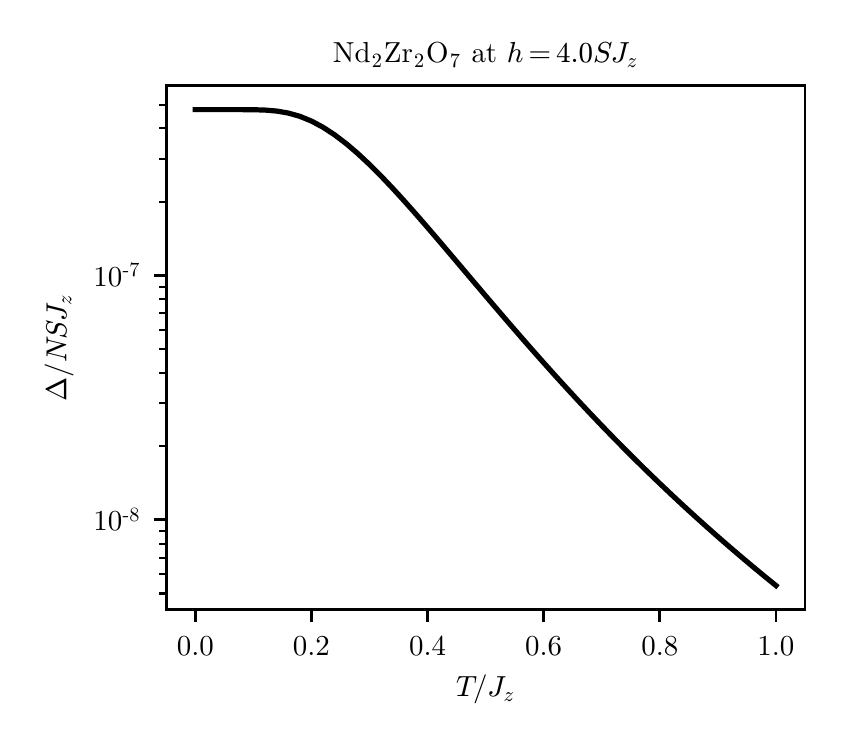}
    \caption{Order-by-disorder as a function of temperature for exchange parameters as estimated for Nd$_2$Zr$_2$O$_7$ (see \autoref{eq:params-Nd2Zr2O7}). The plot shows the gap $\Delta$ between the ferromagnetic chain configuration and the zigzag band. For low temperatures, it approaches the gap in zero point energies while for high temperatures, the gap vanishes.}
    \label{fig:obt-tempterature}
\end{figure}

The second approach treating the finite temperature is considering the gas of non-interacting magnons. For that, we start from the spin wave Hamiltonian in \autoref{app2-lswt-hamiltonian}, which after a Bogoliubov transformation takes the form
\begin{align}
    \mathcal H_2 = \sum_{\vec q, \nu} \omega_\nu(\vec q) \left( n_\nu(\vec q) + \frac{1}{2} \right)
\end{align}
where $n_\nu(\vec q) = \eta_\nu^\dagger(\vec q) \eta_\nu(\vec q)$ is the magnon number operator and the magnon dispersion $\omega_\nu(\vec q)$ is given by the eigenvalues of $\sigma \mathcal X(\vec q)$. Note that these eigenvalue come in pairs $\pm \omega_\nu(\vec q)$ (see \appref{app:lswt} for details). 

The free energy of this system at temperature $T$ is then
\begin{align}
    F_2 &= -T \log(\tr e^{-\beta \mathcal H_2}) \\
    &= \sum_{\vec q, \nu} T\log(1-e^{-\beta\omega_\nu(\vec q)})+\frac{1}{2}\omega_\nu(\vec q)
    \label{eq:F-bose-gas}
\end{align}
where the second term in the sum yields the zero point energy and the first term is a finite-temperature correction. For zero temperature, the above is equal to the zero point energy while for large temperature it becomes independent of the classical ground state configuration $\{\vec S^{(0)}_j\}$. Particularly, for all temperatures the same ground state is favored by fluctuation of the free Bose gas but the gap in free energy vanishes as $T\to\infty$. As an example, we show the free energy gap $\Delta$ between the ferromagnetic chain configuration and the zigzag band as a function of temperature, for exchange parameters as estimated for Nd$_2$Zr$_2$O$_7$, in \autoref{fig:obt-tempterature}.

The vanishing order-by-disorder in the high temperature limit can be understood by expanding the exponential inside the logarithm in \autoref{eq:F-bose-gas}
\begin{align}
    &F_2 =\nonumber \\
    &
    \sum_{\vec q, \nu} T\log(\beta\omega_\nu(\vec q) - \frac{1}{2}(\beta\omega_\nu(\vec q))^2 + \order{\left(\beta\omega_\nu(\vec q)\right)^3})
    %\nonumber \\
    %& 
    %\qquad \qquad
    +\frac{1}{2}\omega_\nu(\vec q) \nonumber\\
    &= \sum_{\vec q, \nu} T\log[\omega_\nu(\vec q)] - T\log(T) + \order{\beta\omega_\nu(\vec q)^2}.
\end{align}
Note that the Eigenvalues of $\sigma \mathcal X(\vec q)$ are exactly the pairs $\pm\omega_\nu(\vec q)$ and the number of eigenvalues $M$ is even such that $\det \sigma \mathcal X(\vec q)$ is positive. Hence, we can write the leading order term of $F_2$ as
\begin{align}
    T \sum_{\vec q, \nu} \log[\omega_\nu(\vec q)] &= \frac{T}{2} \sum_{\vec q} \log \det \sigma \mathcal X(\vec q).
\end{align}
Now, using that $\det \sigma = 1$ and that the quantum spin wave matrix $\mathcal X(\vec q)$ is related to its classical counterpart $\mathfrak X(\vec q)$ (defined in \autoref{eq:app4-low-t-H}) by a unitary transformation [\autoref{eq:x-cq-trafo}], we arrive at
\begin{align}
    T \sum_{\vec q, \nu} \log[\omega_\nu(\vec q)] = \frac{T}{2} \sum_{\vec q} \log \det \mathfrak X(\vec q).
\end{align}
At infinite temperature, we hence recover the purely classical result where, as discussed in the previous section of this appendix, there is no order by disorder.

We note that the discussion of finite-temperature OBD as presented here is not comprehensive as it does not incorporate the possible presence of domain walls at finite temperature. However, since in the presence of anisotropies domain wall excitations, just as spin wave excitations, are gapped, we expect our results to be robust in the low-temperature limit.

\bibliography{references.bib}

\end{document}